\documentclass[reprint,english,preprintnumbers,amsmath,amssymb,aps,prx]{revtex4-2}

\usepackage{amsmath,amssymb}
\usepackage{graphicx}
\usepackage{rotating}
\usepackage{dsfont}
\usepackage{color}
\usepackage[dvips]{epsfig}     
\usepackage{multirow}
\usepackage{breakurl}
\usepackage{babel}
\usepackage[utf8]{inputenc}
\usepackage{slashed}

\usepackage{caption}
\usepackage{subcaption}
\usepackage{charter} 
\usepackage[charter]{mathdesign}

\allowdisplaybreaks

\definecolor{refcol}{RGB}{178,34,34}
\usepackage{microtype}
\usepackage[colorlinks,linkcolor=refcol,citecolor=refcol,urlcolor=refcol]{hyperref}

\graphicspath{{./figures/}}

\newcommand{\tr}{\mathrm{tr}}
\def\det{{\mathrm{det}}}

\def\eq#1{(\ref{#1})}

\def\Eq#1{Eq.~(\ref{#1})}

\def\Fig#1{Fig.~\ref{#1}}

\def\Sec#1{Sec.~\ref{#1}}
\def\App#1{App.~\ref{#1}}

\definecolor{blue}{rgb}{0,0,1}

\definecolor{green}{rgb}{0,1,0}

\definecolor{red}{rgb}{1,0,0}

\begin{document}

\title{Higher topological charge in the QCD vacuum and axion cosmology}

\author{Fabian Rennecke}
\email[E-mail: ]{frennecke@bnl.gov}
\affiliation{Physics Department, Brookhaven National Laboratory,
  Upton, NY 11973, USA}

\begin{abstract}
It is shown that gauge field configurations with higher topological charge modify the structure of the QCD vacuum, which is reflected in its dependence on the CP-violating topological phase $\theta$. To explore this, topological susceptibilities and the production of axion dark matter are studied here. The former characterize the topological charge distribution and are therefore sensitive probes of the topological structure of QCD. The latter depends on the effective potential of axions, which is determined by the $\theta$-dependence of QCD. The production of cold dark matter through the vacuum realignement mechanism of axions can therefore be affected by higher topological charge effects. This is discussed qualitatively in the deconfined phase at high temperatures, where a description based on a dilute gas of instantons with arbitrary topological charge is valid. As a result, topological susceptibilities exhibit a characteristic temperature dependence due to anharmonic modifications of the $\theta$-dependence. Furthermore, multi-instanton effects give rise to a topological mechanism to increase the amount of axion dark matter.
\end{abstract}

\maketitle


\section{Introduction}\label{sec:intro}

The vacuum structure of interacting quantum field theories is inherently intriguing. In quantum chromodynamics (QCD) and other gauge theories, it is well known that this structure crucially depends on topological gauge field configurations \cite{tHooft:1976rip, tHooft:1976snw, tHooft:1986ooh, Jackiw:1976pf, Callan:1976je}. 
A vacuum state can be characterized by an integer, the topological charge, and topological gauge fields describe tunneling processes between topologically distinct realizations of the vacuum. The true vacuum is therefore a superposition of all possible topological vacuum configurations. It is characterized by a free parameter, the CP-violating topological phase $\theta$, acting as a source for topological charge correlations.

The topological nature of the QCD vacuum has important phenomenological consequences. The axial anomaly in QCD is realized through topologically nontrivial fluctuations. They generate anomalous quarks correlations which explicitly break $U(1)_A$ and impact properties of hadrons \cite{Pisarski:2019upw}. Most prominently, the large mass of the 
$\eta^\prime$ meson is generated by topological effects \cite{tHooft:1976rip, tHooft:1976snw, tHooft:1986ooh, Witten:1979vv, Veneziano:1979ec}. The fate of the axial anomaly at finite temperature also determines the order of the QCD phase transition in the limit of massless up and down quarks \cite{Pisarski:1983ms, Resch:2017vjs}.
Furthermore, a nonzero $\theta$ introduces CP-violating strong interactions. These would manifest in a nonvanishing electric dipole moment of the neutron, $d_n$. Recent measurements give a stringent bound of $|d_n| < 1.8 \times 10^{-26} e\, \text{cm}$ \cite{Abel:2020gbr}, resulting in $\theta \lesssim 10^{-10}$ \cite{Crewther:1979pi, Pospelov:1999mv}. This strongly suggests that there is no strong CP-violation. The question about the existence and nature of a physical mechanism to enforce $\theta = 0$ remains to be answered.

One resolution of this strong-CP problem has been suggested by Peccei and Quinn (PQ) \cite{Peccei:1977hh}. It involves the introduction of a new complex scalar field, which has a global chiral $U(1)_\text{PQ}$ symmetry. The PQ symmetry is broken both explicitly, through the axial anomaly, and spontaneously. The resulting pseudo-Goldstone boson, the axion $a(x)$ \cite{Weinberg:1977ma, Wilczek:1977pj}, couples to the gauge fields in exactly the same way as the $\theta$-parameter, $\sim (a/f_a + \theta)\, \tr\, F \tilde F$, where $F$ and $\tilde F$ are the gluon field strength and its dual, and $f_a$ is the axion decay constant. This effectively promotes 
$\theta$ to a dynamical field, which has a physical value given by the minimum of its effective potential. Owing to the anomaly, the effective potential of axions is determined by topological gauge field configurations, and its minimum is at zero, thus resolving the strong CP-problem dynamically. Furthermore, the anomalous axion mass is very small and, in order to be consistent with observational bounds, axions couple very weakly to ordinary matter. This makes axions attractive dark matter candidates \cite{Sikivie:2006ni, Marsh:2015xka, DiLuzio:2020wdo}. Cold axions can be produced non-thermally through a field relaxation process known as vacuum realignement \cite{Preskill:1982cy, Abbott:1982af, Dine:1982ah, Turner:1985si}. The axion relaxes from a, possibly large, initial value to its vacuum expectation value at zero, thereby probing the global structure of the $\theta$-vacuum. The production of cold dark matter through the vacuum realignment mechanism of axions therefore is particularly sensitive to the topological structure of the QCD vacuum.

While the nature of topological field configurations in the confined phase is still unsettled, at large temperatures where the gauge coupling is small, a semi-classical analysis is valid \cite{Pisarski:1980md, Gross:1980br} and these field configurations can be described by instantons \cite{Belavin:1975fg}, see \cite{Vainshtein:1981wh, Schafer:1996wv, Diakonov:2002fq} for reviews. Finite temperature is crucial for a self-consistent treatment, since it effectively cuts off large-scale instantons. This is clearly shown by numerous first-principles studies of QCD on the lattice, both with dynamical quarks \cite{Bonati:2015vqz, Petreczky:2016vrs, Borsanyi:2016ksw, Burger:2018fvb, Jahn:2020oqf, Lombardo:2020bvn}, and in the quenched limit \cite{Bonati:2013tt, Borsanyi:2015cka, Xiong:2015dya, Berkowitz:2015aua}. It has been demonstrated that the temperature dependence of the lowest topological susceptibility $\chi_2$, which measures the variance of the topological charge distribution, is in very good agreement with the corresponding prediction from a dilute gas of instantons for $T \gtrsim 2.5 T_c$, where $T_c$ is the pseudocritical temperature of the chiral phase transition. One can therefore assume that the topological structure of the QCD vacuum is described by instantons at large temperatures. This is of relevance also for axion physics, since, for example, the axion mass is proportional to the topological susceptibility, $\chi_2 = f_a^2 m_a^2$.

Conventionally, only single-instantons, i.e.\ instantons with unit topological charge, are taken into account in the description of the topological structure of the QCD vacuum at large temperatures. The reason is that classically the action of an instanton with topological charge $Q$ in a dilute gas is $\sim e^{- 8 \pi^2 |Q|/g^2}$. Thus, in the limit of vanishing gauge coupling $g$, multi-instantons with $|Q|>1$ are subject to a large exponential suppression.
Furthermore, the moduli space of multi-instantons coincides with that of $|Q|$ independent single-instantons, which are each characterized by a position $z_i$, a size $\rho_i$ and an orientation in the gauge group $U_i$, with $i = 1,\dots, |Q|$ \cite{Bernard:1977nr}.
Yet, multi-instantons are distinct self-dual gauge field configurations which, in general, cannot be interpreted as simple superpositions of single-instantons \cite{Atiyah:1978ri, Christ:1978jy}. 
Since quantum corrections lead to an increasingly strong coupling towards lower energies, multi-instantons could give relevant corrections to the leading single-instanton contribution. Furthermore, it has been shown in \cite{Pisarski:2019upw} that there are effects that are related uniquely to higher topological charge: higher order anomalous quark interactions are generated only by multi-instantons, which generalizes the classic analysis for single-instantons \cite{tHooft:1976rip, tHooft:1976snw}.
Thus, one has to assume that the topological field configurations at large temperatures are instantons of arbitrary topological charge.

Given this motivation, the effects of multi-instantons on the vacuum structure of QCD are explored in this work. It is assumed that at sufficiently high temperature in the deconfined phase, the topological structure of QCD is described by a dilute gas of instantons of arbitrary topological charge. This leads to a modification of the known $\theta$-dependence of the QCD vacuum. It is reflected in the distribution of topological charge, which is probed by topological susceptibilities. In addition, as outlined above, axion cosmology is an interesting application to showcase the effect of higher topological charge on the QCD vacuum. To this end, the production of cold axion dark matter via the vacuum realignment mechanism is investigated here as well.

A semi-classical description of multi-instantons requires knowledge of the partition function of QCD in their presence. At next-to-leading order in the saddle point approximation, the complete partition function is known for single-instantons at finite temperature \cite{tHooft:1976rip, tHooft:1976snw, Pisarski:1980md, Gross:1980br}. For multi-instantons, it is only known in certain limits, see \cite{Osborn:1981yf, Dorey:2002ik} for reviews. The exact multi-instanton solutions \cite{Atiyah:1978ri} can be expanded systematically if the size parameters $\rho_i$ are small against the separation
$|R_{ij}| = |z_i-z_j|$ for $i\neq j$ and $i,j = 1,\dots, |Q|$ \cite{Christ:1978jy}. At leading order in this limit of small constituent-instantons (SCI), the partition function of QCD in the background of a multi-instanton can be computed solely based on the knowledge of the single-instanton solution. At next-to-leading order, correlations between constituent-instantons need to be taken into account \cite{Brown:1978yj, Bernard:1978ea, Pisarski:2019upw}. The resulting genuine multi-instanton processes are studied in a qualitative manner in the present work. Due to the large suppression of the instanton density in the presence of dynamical quarks, it is expected that, at least in the large temperature regime, multi-instanton effects are most pronounced in quenched QCD, i.e.\ without dynamical quarks. Hence, quenched QCD provides a good laboratory to understand higher topological charge effects on the QCD vacuum.

This work is organized as follows: The $\theta$-dependence in a dilute gas of multi-instantons is derived in \Sec{sec:dil}. This requires knowledge of the multi-instanton contribution to the partition function, which is derived in detail in the SCI limit in \Sec{sec:ZQSCI}. The result is used to study topological susceptibilities in \Sec{sec:sus}. Their temperature dependence is evaluated numerically for quenched QCD in \Sec{sec:YMsus} and discussed for QCD in \Sec{sec:QCDsus}. The focus there is on systematically investigating correction to the topological susceptibilities as instantons with increasing topological charge are included. The impact of multi-instantons on axion cosmology is studied in \Sec{sec:axi}. First, the resulting axion effective potential and the production of axions via the vacuum realignment mechanism are discussed in Secs.\ \ref{sec:axipot} and \ref{sec:axicos}. In \Sec{sec:YMaxi} axion production is studied numerically in the quenched approximation. Finally, a critical discussion of the approximations used here can be found in \Sec{sec:disc}, and a summary of the results in \Sec{sec:sum}. Further technical details are given in the appendices.

\section{$\theta$-dependence from a dilute gas of multi-instantons}\label{sec:dil}

The starting point is a modification of the old story of the $\theta$-dependence of the QCD vacuum. In the context of the axial anomaly and the $U(1)_A$ problem in QCD, it has been realized that the QCD vacuum has more complicated structure which is related to the existence of topological gauge field configurations \cite{tHooft:1986ooh, Jackiw:1976pf, Callan:1976je}. Physical states can therefore be grouped into homotopy classes of field configurations with a given topological charge (or winding number) $| n \rangle$,
\begin{align}\label{eq:topch}
n = -\frac{1}{16 \pi^2}  \int\!\! d^4x\, \tr\, F_{\mu\nu} \tilde F_{\mu\nu} \,\in \mathbb{Z}\,,
\end{align}
where $F_{\mu\nu} = [D_\mu , D_\nu]$ is the field strength tensor with $D_\mu = \partial_\mu + A_\mu$ the covariant derivative.  $\tilde F_{\mu\nu} = \frac{1}{2}\epsilon^{\mu\nu\rho\sigma} F_{\rho\sigma}$ is the corresponding dual field strength. Instantons of topological charge $Q$
can be interpreted as tunneling from state $| n \rangle$ to state 
\mbox{$| n + Q  \rangle$}, with the tunneling amplitude related to the exponential of the classical instanton action, $e^{- 8 \pi^2 |Q|/g^2}$. Due to such tunneling processes, $| n \rangle$ cannot describe the vacuum state in a unique way. Furthermore, states characterized by different winding numbers are related to each other by large gauge transformations. The true vacuum state, consistent with gauge symmetry, locality and cluster decomposition is the $\theta$-vacuum,
\begin{align}
|\theta\rangle = \sum_{n=-\infty}^{+\infty} e^{-i n \theta}\, | n \rangle.
\end{align}
An important property of this state is that the value of $\theta$ cannot be changed by a gauge invariant operation. Hence, QCD falls into superselection sectors, where each $\theta$ labels a different theory.

The vacuum amplitude between in- and out-states at $t = -\infty$ and $+\infty$, denoted by $| \theta \rangle_-$ and ${}_+{\langle \theta |}$ respectively, is then given by
\begin{align}\label{eq:vacamp1}
\begin{split}
{}_+\langle \theta | \theta \rangle_- &= \sum_n \sum_m e^{i m \theta} e^{-i n \theta} {}_+\langle m | n \rangle_-\\
&= \sum_\nu e^{i \nu\, \theta} \sum_n {}_+\langle n+ \nu | n \rangle_- \,,
\end{split}
\end{align}
with $\nu = m-n$. The part after the exponential in the second line of this equation describes the amplitude where in- and out-state differ by a net-topological charge of $\nu$. Using \Eq{eq:topch}, the $\theta$-dependent generating functional $\mathcal{Z}[\theta] \equiv {}_+\langle \theta | \theta \rangle_- $ can be represented by the Euclidean path integral
\begin{align}\label{eq:zthet1}
\begin{split}
\mathcal{Z}[\theta]= \sum_\nu \int\! &\mathcal{D}\Phi\, e^{-\, S[\Phi]\, - \frac{i \theta}{16\pi^2}  \int\!\! d^4x\, \tr\, F_{\mu\nu} \tilde F_{\mu\nu}}\\
&\times\delta \bigg(\! \nu + \frac{1}{16\pi^2}  \int\!\! d^4x\, \tr\, F_{\mu\nu} \tilde F_{\mu\nu} \bigg)\,,
\end{split}
\end{align}
where the multi-field $\Phi = (A_\mu, q, \bar q, c, \bar c)$ contains the gluon, quark and ghost fields and $S[\Phi]$ is the action of gauge-fixed Euclidean QCD. There can also be source terms, but they are suppressed here for the sake of brevity.

So far, no reference to the nature of the topological field configurations has been made. With the corresponding remarks in Sects.\ \ref{sec:intro} and \ref{sec:disc} in mind, assume that these configurations are given by instantons. 
In addition, assume that the instantons are dilute, i.e.\ the spacetime distance between each instanton is large against their effective size. It has been demonstrated by numerous studies of QCD and Yang-Mills theory on the lattice that these assumptions are justified at temperatures above about $2.5 T_c$ \cite{Bonati:2015vqz, Petreczky:2016vrs, Borsanyi:2016ksw, Burger:2018fvb, Jahn:2020oqf, Lombardo:2020bvn, Bonati:2013tt, Borsanyi:2015cka, Xiong:2015dya, Berkowitz:2015aua}. Note that, while the overall power of the topological susceptibility with respect to $T$ measured on the lattice agrees very well with the predictions of the dilute instanton gas, the prefactor is off. This can likely be attributed to missing higher loop corrections, which can be substantial in the hot QCD medium, see, e.g., the discussion in \cite{Petreczky:2016vrs}.

What is new here is that the contributions of multi-instantons with arbitrary topological charge $Q$ are also taken into account.
Since their contributions to the path integral are proportional to $e^{-8 \pi^2 |Q|/g^2}$ at weak coupling, single-instantons clearly dominate in the semi-classical regime. For a comprehensive discussion of the conventional dilute instanton gas, see \cite{Coleman:1978ae}. Yet, as discussed above, there is no reason to assume that field configurations with higher topological charge are not present. Since the $\theta$-dependence is directly linked to the topological structure of the vacuum, it is conceivable that it is affected by multi-instantons. Furthermore, it has been shown in \cite{Pisarski:2019upw} that there are effects that can be related uniquely to multi-instantons.

\begin{figure}[t]
\centering
\includegraphics[width=.37\textwidth]{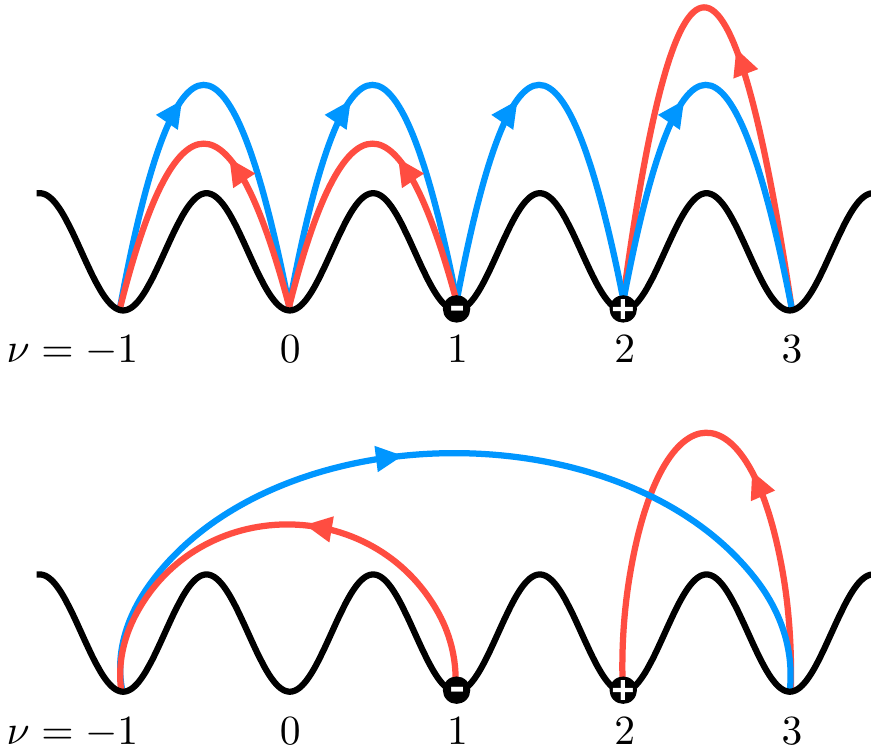}
\caption{Illustration of possible instanton configurations contributing to the vacuum amplitude ${}_+\langle 2 | 1 \rangle_-$, which is part of the complete $\theta$-vacuum amplitude in \Eq{eq:vacamp1}. The black sinosodial represents the vacuum, where each minimum corresponds to a different winding number $\nu$. In- and out-states are marked by - and +, respectively. The blue lines show the vacuum transition due to instantons and the red lines the transitions due to anti-instantons. The size of the transition, i.e.\ the difference in winding from the starting point to the endpoint of an instanton, correspond to its topological charge. The upper figure illustrates a contribution where only single-instantons are considered. Exclusively configurations of such type are taken into account in the conventional dilute instanton gas. The lower figure shows a similar contribution, but including multi-instantons. The present analysis takes all possible configurations, including single- and multi-instantons, into account.}
\label{fig:tun}
\end{figure}

There are various possibilities to realize the vacuum amplitude in \Eq{eq:zthet1} for a given $\nu$ in a dilute gas of multi-instantons. If only single-instantons are taken into account, all configurations where $n_1$ instantons and $\bar n_1$ anti-instantons are distributed in spacetime in a dilute way are allowed, as long as $\nu = n_1 - \bar n_1$ is fulfilled. This can be generalized straightforwardly for multi-instantons: One may `sprinkle' spacetime with all possible combinations of $n_Q$ $Q$-instantons, i.e.\ instantons with topological charge $Q$, and $\bar n_Q$ $Q$--anti-instantons, provided $\nu = \sum_Q Q (n_Q - \bar n_Q)$ holds. This is illustrated in \Fig{fig:tun}. If not otherwise specified, $Q>0$ is used from now on. The difference between positive and negative topological charge is made by referring to instantons and anti-instantons, respectively.

The contribution of one multi-instanton $A_\mu^{(Q)}$ to the vacuum amplitude is given by the path integral in the background of $A_\mu^{(Q)}$,
\begin{align}\label{eq:piq}
\begin{split}
e^{i \theta Q}\, Z_Q = e^{i \theta Q} \int\! \mathcal{D}\Phi\, e^{-S[\Phi+\bar\Phi^{(Q)}]}\,,
\end{split}
\end{align}
where $\Phi$ is the fluctuating multi-field and $\bar\Phi^{(Q)} = (A_\mu^{(Q)}, 0, 0, 0, 0)$ is the background field. The underlying assumption here is that the $Q$-instanton is the only topological field configuration. This is certainly guaranteed in a saddle point approximation about the background field, where only small fluctuations around the $Q$-instanton background are assumed. In the dilute limit, the $\theta$-vacuum amplitude $\mathcal{Z}[\theta]$ is then completely described by a statistical ensemble of all possible combinations of $Z_Q$'s.

As mentioned above, a multi-instanton $A_\mu^{(Q)}$ is described by free parameters known as collective coordinates. They can be interpreted as positions $z_i$, sizes $\rho_i$ and orientations $U_i \in SU(N_c)$ of $Q$ constituent-instantons with unit topological charge \cite{Bernard:1977nr, Atiyah:1978ri, Christ:1978jy}. As discussed in more detail below, the partition function $Z_Q$ can formally be written as an integral over these parameters.
It has been shown in Refs.\ \cite{Brown:1978yj, Bernard:1978ea, Pisarski:2019upw} that if the relative distances between the constituent-instantons, $|R_{ij}|$, are small against their sizes, $\rho_i$, the path integral factorizes into individual contributions of the constituent-instantons. The underlying reason is that the interaction between instantons is very short-ranged. At this leading order in the SCI limit, the multi-instanton itself can be viewed as a simple superposition of single-instantons. Furthermore, large-scale instantons are suppressed at finite temperature \cite{Pisarski:1980md, Gross:1980br}.
This is the foundation for using a dilute gas of single-instantons to describe the topological structure of QCD at large temperatures.

Genuine multi-instanton processes can only contribute to \Eq{eq:piq} at large temperature if the constituent-instantons are close together. Since $Z_Q$ involves integrations over all possible constituent-instanton positions, there is a hierarchy of instanton contributions: If all constituent-instantons are far apart, only single-instantons contribute; if two constituent-instantons are close together, genuine 2-instanton processes contribute; and so forth until $Q$ constituent-instantons are close together, which gives a genuine $Q$-instanton contribution. Since instanton interactions are short ranged, $n$-instantons with $1\leq n \leq Q$, contribute to $Z_Q$ roughly proportional to a spacetime-volume factor $\mathcal{V}^{Q- n +1}$. One factor of volume $\mathcal{V}$ always arises since only the relative positions of constituent-instantons matter.

$Z_Q$ can therefore be decomposed into a part that is only due to $q$-instanton processes $Z_Q^{(q)}$ with $q = 1,\dots, Q-1$, and a genuine Q-instanton contribution $\Delta Z_Q$,
\begin{align}\label{eq:ZQdec}
Z_Q = \sum_{q=1}^{Q-1} Z_Q^{(q)} + \Delta Z_Q\,.
\end{align}
For ease of notation, $Z_1^{(1)} = 0$ is used for $Q=1$, such that $Z_1 = \Delta Z_1$.
This forms the basis for a dilute gas of multi-instantons, which incorporates subleading corrections to the conventional dilute instanton gas. The distinction between $q$-instanton and genuine $Q$-instanton processes in $Z_Q$ is necessary to avoid double-counting in the total vacuum amplitude of the dilute gas. Since in the classical limit the contribution of $Q$ single-instantons to the vacuum amplitude is exactly the same as of a single $Q$-instanton, genuine multi-instanton contributions are only due to quantum corrections.

The dilute multi-instanton gas is a statistical ensemble of single- and multi-instanton processes $\Delta Z_Q$.
The resulting free energy density, $F = -\frac{1}{\mathcal{V}} \ln \mathcal{Z}$, in the presence of a $\theta$-term is then:
\begin{align}\label{eq:Fgen}
\nonumber
F(\theta) &= - \frac{1}{\mathcal{V}} \ln \sum_\nu \Bigg\{ \prod_Q \sum_{n_Q = 0}^\infty \sum_{\bar n_Q = 0}^\infty \frac{1}{n_Q! \bar n_Q!}\, \Delta Z_Q^{n_Q + \bar n_Q}\, \\  \nonumber
&\quad \times e^{i \theta Q (n_Q - \bar n_Q)}\, \delta^\nu_{\sum_Q Q (n_Q - \bar n_Q)}\Bigg\}\\
&= -  \frac{2}{\mathcal{V}} \sum_Q \Delta Z_Q \cos(Q \theta)\,,
\end{align}
where $\mathcal{V} = V/T$ is the Euclidean spacetime volume. The $\theta$-vacuum entails a sum over all $\nu$, and the Kronecker delta at the end of the second line enforces net winding $\nu$ for the sum over all possible combinations of (multi-) instantons. Since the sum over all possible multi-instanton configurations yields all possible net windings, one can drop both the sum over $\nu$ and the Kronecker delta.
For reasons that become clear below, it is convenient to define the $\theta$-dependent part of the free energy,
\begin{align}\label{eq:DF}
\begin{split}
\Delta F(\theta) &= F(\theta) - F(0)
=  \frac{2}{\mathcal{V}} \sum_Q \Delta Z_Q \Big[1- \cos(Q \theta)\Big]\,.
\end{split}
\end{align}
Hence, while the $\theta$-dependence generated only by single-single instantons is simply $\sim \cos \theta$, the inclusion of multi-instantons reveals a much richer structure. This result can be viewed as an expansion of the free energy in terms of multi-instanton contributions. It is worth emphasizing that, since $Q \in \mathbb{Z}$, the $2\pi$-periodicity of the $\theta$-vacuum is guaranteed also here. The corrections due to multi-instantons are overtones to the single-instanton contribution, which sets the fundamental frequency.

\subsection{Multi-instanton contribution from small constituent-instantons}\label{sec:ZQSCI}

To evaluate the free energy explicitly, knowledge of $\Delta Z_Q$ is required.
As discussed above, only certain limits are known, one of them being the SCI limit. 
It uses that the exact multi-instanton solution can be expanded systematically in powers of $\rho/|R|$ \cite{Christ:1978jy}. To leading order in this expansion, the $Q$-instanton can be interpreted as a superposition of $Q$ constituent-instantons with unit topological charge. 
Remarkably, to order $\rho^2/|R|^2$, the path integral of QCD factorizes into the contributions of the individual constituent-instantons with $Q = 1$ \cite{Brown:1978yj, Bernard:1978ea, Pisarski:2019upw}, which yields the first term of \Eq{eq:ZQdec},
\begin{align}\label{eq:ZQLO}
Z_Q^{(1)} = \frac{1}{Q!} Z_1^Q \,.
\end{align}
The combinatorial prefactor $1/Q!$ arises because the constituent-instantons can be treated as identical particles.
$Z_1$ can be expressed in terms of the single-instanton density $n_1$:
\begin{align}\label{eq:Z1}
Z_1 &= \int \! d^4z\, d\rho\, n_1(\rho, T)
\equiv \mathcal{V} \bar Z_1\,.
\end{align}
The single-instanton density only depends on the instanton size and the temperature. The integration over the instanton location $z$ gives a factor of spacetime volume $\mathcal{V}$. $n_1$ has been computed to next-to-leading order in the saddle point approximation in \cite{tHooft:1976rip, tHooft:1976snw, Bernard:1979qt, Morris:1984zi, Pisarski:1980md, Gross:1980br, Altes:2014bwa}, see \App{app:id} for details.
Provided that the constituent-instantons are sufficiently far apart, any vacuum amplitude with net topological charge $Q$ can be described by a superposition of single-instanton processes only. This is exemplified in the top figure of \Fig{fig:tun}. The $Q=1$ contribution in \Eq{eq:Fgen} accounts for all these processes.

For genuine multi-instanton processes, as shown in the bottom figure of \Fig{fig:tun}, correlations between constituent-instantons need to be taken into account. So while \Eq{eq:ZQLO} holds in case the constituent-instantons are far apart, distinct multi-instanton contributions arise when they overlap to some extent. For the gauge fields themselves, these corrections are of order $\rho^4/|R|^4$ \cite{Pisarski:2019upw, Christ:1978jy}. However, due to the existence of quark zero modes in the presence of instantons, the first correction to the leading-order SCI-limit arises at order $\rho^3/|R|^3$ \cite{Pisarski:2019upw}. In Ref.\ \cite{Pisarski:2019upw} the correlation of constituent-instantons has been computed explicitly for $Q = 2$. Here, this to generalized to $Q \geq 2$.

The generating functional in the background of a $Q$-instanton is defined in \Eq{eq:piq}. It is instructive to write this in terms of the QCD action in the chiral limit, i.e.\ with vanishing quark mass matrix $M_q$, and an explicit bilinear quark term,
\begin{align}\label{eq:ZQgen}
\begin{split}
Z_Q[J] = \int\! \mathcal{D}\Phi\,\exp \Bigg\{&-S\big[\Phi + \bar\Phi^{(Q)}\big]\Big|_{M_q=0}\\
&+ \int\!d^4x\, \bar\psi(x)\, J(x)\, \psi(x)  \Bigg\}\,,
\end{split}
\end{align} 
where the source $J$ can be set equal to $M_q$ to recover the original action. For the following computation to be valid also in the chiral limit at large temperature, it is useful to keep a more general source $J$, which may have nontrivial flavor and spinor structure. To next-to-leading order in the saddle point approximation, $Z_Q[J]$ can be written in term of the functional determinants of the fluctuating quark, gluon and ghost fields. However, the instanton collective coordinates arise from translations, dilatations and global gauge rotations which are symmetries of the system, but yield inequivalent instanton solutions. Thus, the collective coordinates correspond to zero mode directions of the gauge fields. Fluctuations around these directions cannot be assumed to be small and need to be treated exactly. The functional integral over the zero modes is replaced by and integral over the collective coordinates, while the nonzero-modes can be computed in the ordinary fashion \cite{tHooft:1976snw, Osborn:1981yf, Dorey:2002ik}.

In addition to the zero modes of gluons and ghost, due to the axial anomaly, the Dirac operator has $N_f Q$ left-handed (right-handed) quark zero modes in the presence of a $Q$-instanton ($Q$--anti-instanton) \cite{tHooft:1976rip, Grossman:1978, Brown:1977bj, Corrigan:1978ce, Osborn:1978rn},
\begin{align}\label{eq:AQDE}
\gamma_\mu \big(\partial_\mu + A_\mu^{(Q)} \big) \psi^{(Q)} = 0\,.
\end{align}
The quark zero modes are functions of the collective coordinates. The resulting generating functional can be expressed in terms of the multi-instanton density $n_Q$,
\begin{align} \label{eq:ZQnQ}
Z_Q[J] &=  \int\! \bigg( N \prod_{i=1}^Q d^4z_i\, d\rho_i\, dU_i \bigg)\, n_Q\big(\{z_i,\rho_i,U_i\}\big)\\ \nonumber
&= \int\! \bigg( N \prod_{i=1}^Q d^4z_i\, d\rho_i\, dU_i \bigg)\, \bar n_Q\big(\{z_i,\rho_i,U_i\}\big)\, \det{}_0^{(Q)}(J)\,,
\end{align}
where $N$ is the normalization of the collective coordinate measure. In the second line, the determinant over the zero modes of quarks, $\det{}_0^{(Q)}(J)$, has been separated from the instanton density for later convenience. It is assumed that $J$ is only a small perturbation to the Dirac operator, such that it doesn't affect the nonzero-modes of the quarks \footnote{This is guaranteed if the source is identified with the mass matrix of the light quarks. There are quantitative modifications for heavier quarks \cite{Dunne:2004sx, Dunne:2005te}, but they are irrelevant here.}. 

To systematically compute $Z_Q[J]$ in the SCI limit, the ADHM construction \cite{Atiyah:1978ri} is used. The following explicit construction is based on \cite{Christ:1978jy} for $N_c = 2$. The straightforward generalization to any number of colors is done in the end. The most general $Q$-instanton can be constructed algebraically from two matrices $M$ and $N$ that obey the simple constraints listed below.
$M$ is a $(Q+1) \times Q$ matrix with quaternionic matrix elements, which can be represented as $2\times2$ matrices
\begin{align}\label{eq:qmat}
M_{ab} = \alpha_\mu M^\mu_{ab}\,.
\end{align}
$M^\mu_{ab}$ are real coefficients and $\alpha_\mu$ are the basis quaternions,
\begin{align}
\alpha_\mu = (\mathds{1}_2, -i \vec{\sigma})_\mu\,,
\end{align}
where $\vec{\sigma}$ are the Pauli matrices.
For the construction of $Q$--anti-instantons, one simply has to replace $\alpha_\mu$ by 
\begin{align}
\bar\alpha_\mu = (\mathds{1}_2, i \vec{\sigma})_\mu\,.
\end{align}
$M(x)$ is chosen to be linear in Euclidean spacetime, $x = \alpha_\mu x^\mu$,
\begin{align}\label{eq:MBCx}
M(x) = B - C x\,.
\end{align}
$B$ and $C$ are constant $(Q+1)\times Q$ matrices of rank $Q$. 
$M$ is required to obey the reality condition
\begin{align}\label{eq:ADHM1}
M^\dagger(x) M(x) = R(x)\,,
\end{align}
with the real, invertible, quaternionic $Q\times Q$ matrix $R$. Each matrix element of a real quaternionic matrix is proportional to $\alpha_0 = \mathds{1}_2$. The quaternionic conjugate is defined as
\begin{align}
\big(M^\dagger\big)_{ab}^0 = M_{ba}^0\,,\qquad \big(M^\dagger\big)_{ab}^i = - M_{ba}^i\,.
\end{align}
$N(x)$ is a quaternionic $(Q+1)$ column vector that obeys the two constraints
\begin{align}\label{eq:ADHM23}
\begin{split}
N^\dagger(x) M(x) &= 0\,,\\
N^\dagger(x) N(x) &= \mathds{1}_2\,.
\end{split}
\end{align}
Remarkably, Eqs.\ \eq{eq:ADHM1} and \eq{eq:ADHM23} determine $M$ and $N$ up to $8 Q$ free parameters, which corresponds exactly to the number of instanton collective coordinates for $N_c = 2$. The exact $Q$-instanton is then simply given by
\begin{align}\label{eq:ADHMA}
A_\mu^{(Q)} = N^\dagger(x) \partial_\mu N(x)\,.
\end{align}
To find $M$ and $N$, one uses that Eqs.\ \eq{eq:ADHM1} and \eq{eq:ADHM23} allow for the transformations $M \rightarrow S M T$ and $N\rightarrow S N$, with an unitary $(Q+1) \times (Q+1)$ matrix $S$ and an invertible, real $Q\times Q$ matrix T, without changing the instanton solution. From Eqs.\ \eq{eq:MBCx} and \eq{eq:ADHM1} follows that $C^\dagger C$ is a symmetric, real, invertible matrix. $T$ can therefore be chosen such that $C^\dagger C$ is congruent to $\mathds{1}_{Q} \alpha_0$. Furthermore, the first row of $S$ can be chosen such that the first row of $C$ vanishes, 
and the other matrix elements can be defined as $S_{(i+1)j} = C^\dagger_{i j}$ with $i= 1.\dots,Q$ and $j= 1.\dots,Q+1$.
After these transformations, $B$ and $C$ in \Eq{eq:MBCx} assume the general form
\begin{align}\label{eq:BC}
\begin{split}
B = 
\begin{pmatrix}
v \\
\bar b
\end{pmatrix}\,, \qquad
C = 
\begin{pmatrix}
0 \\
\mathds{1}_{Q} \alpha_0
\end{pmatrix}\,,
\end{split}
\end{align}
where $v = (q_1,\dots, q_Q)$ is a row vector of arbitrary quaternions $q_i$ and 
\begin{align}
\bar b_{ij} = \delta_{ij} z_i + b_{ij}\,,
\end{align}
is a $Q\times Q$ matrix with arbitrary quaternions $z_i$ and $b_{ij}$. The diagonal elements of $\bar b$ are chosen to be given by $z_i$ and $b_{ii} = 0$ for all $i=1,\dots,Q$. 
$\mathds{1}_{2Q}$ is the real, quaternionic $Q\times Q$ unit matrix.
It follows from \Eq{eq:ADHM1} that the matrix $b$ is symmetric, $b_{ij} = b_{ji}$, and from $2 {\rm Im}\, R_{ij} = R_{ij} - R_{ij}^* = 0$ follows
\begin{align}\label{eq:bijgen}
\begin{split}
&2 (z_i - z_j)^* b_{ij} - 2 {\rm Re\big[(z_i - z_j)^* b_{ij}}\big]\\
&\quad =( q_j^* q_i - q_i^* q_j )+ \sum_{k=1}^Q (b_{kj}^* b_{ki} - b_{ki}^* b_{kj})\,.
\end{split}
\end{align}
$^*$ denotes the quaternionic conjugate, which corresponds to the conjugate transpose of the matrix representation in \Eq{eq:qmat}.
Solving this equation is difficult in general. If the $q_i$ are assumed to be real, then \Eq{eq:bijgen} is solved by $b_{ij} = 0$. As shown in \cite{Christ:1978jy} (and below), this leads to 't Hooft's multi-instanton solution, where all constituent instantons have the same orientation in the gauge group.

More general, self-consistent solutions can be constructed by first noting that if $b_{ij}$ is chosen such that ${\rm Re\big[(z_i - z_j)^* b_{ij}}\big] = 0$, then \Eq{eq:bijgen} can be written as
\begin{align}\label{eq:bijspec}
b_{ij} = \frac{1}{2} \frac{z_i - z_j}{|z_i - z_j|^2} \bigg[\! q_j^* q_i - q_i^* q_j+\! \sum_{k=1}^Q \big(b_{kj}^* b_{ki} - b_{ki}^* b_{kj} \big)\!\bigg]\,.
\end{align}
This equation can be solved systematically in the SCI limit. To this end, write the quaternion $q_i$ in terms of its modulus and its phase,
\begin{align}
q_i = \rho_i U_i\,,
\end{align}
where $\rho_i = \sqrt{q_i^* q_i}$ can readily be interpreted as the size of the $i$-th constituent-instanton and $U_i \in SU(2)$ as its gauge group orientation. In the SCI limit $\rho_i$ is assumed to be small against $|R_{ij}| = |z_i - z_j|$. Hence, one can replace $q_i$ by $\zeta q_i$, with $|q_i|$ held fixed, and expand in powers of the small parameter $\zeta$. Then, the first nonvanishing contribution to $b_{ij}$ in the expansion of \Eq{eq:bijspec} is of order $\zeta^2$,
\begin{align}
b_{ij} = \frac{1}{2} \frac{z_i - z_j}{|z_i - z_j|^2} \big( q_j^* q_i - q_i^* q_j \big) + \mathcal{O}(\zeta^4)\,.
\end{align}
Solutions to \Eq{eq:bijspec} to any power in $\zeta$ can be constructed from this solution by iteration. For small constituent-instantons, the resulting series is guaranteed to be convergent. 
To order $\zeta$, one can therefore set $b_{ij} = 0$ even for arbitrary quaternions $q_i$. This is sufficient for the present purposes. The resulting matrix $M$ is
\begin{align}\label{eq:Msmall}
M=
\begin{pmatrix}
q_1 & \cdots & q_Q\\
z_1-x & \cdots & 0 \\
\vdots & \ddots & \vdots\\
0 & \cdots & z_Q - x
\end{pmatrix} + \mathcal{O}(\zeta^2)\,.
\end{align}
With this, \Eq{eq:ADHM23} can be solved easily,
\begin{align}\label{eq:Nsmall}
N(x) = \frac{1}{\sqrt{\xi_0}}
\begin{pmatrix}
u \\
\frac{x-z_1}{(x-z_1)^2}\, q_1^*\cdot u \\
\vdots \\
\frac{x-z_Q}{(x-z_Q)^2}\, q_Q^*\cdot u
\end{pmatrix} + \mathcal{O}(\zeta^2)\,,
\end{align}
where the normalization is determined by
\begin{align}
\xi_0(x) =1+ \sum_{i= 1}^{Q} \frac{\rho_i^2}{(x - z_i)^2}\,.
\end{align}
$u$ is an arbitrary unit quaternion. Different choices for $u$ give gauge-equivalent instanton fields. The simplest choice is $u = \alpha_0$, which corresponds to the singular gauge. Indeed, plugging this into \Eq{eq:ADHMA} yields the $Q$-instanton to order $\zeta^2$ in the SCI limit,
\begin{align}\label{eq:AQsmall}
A_\mu^{(Q)}(x) = \frac{1}{\xi_0(x) } \sum_{i=1}^{Q} U_i \bar\sigma^{\mu\nu} U_i^\dagger\, \rho_i^2\, \frac{ (x-z_i)_\nu}{|x-z_i|^4} + \mathcal{O}(\zeta^4)\,.
\end{align}
The anti-selfdual matrix $\bar\sigma^{\mu\nu}$ is defined as
\begin{align}
\bar\sigma^{\mu\nu} = \frac{1}{2} \big(\alpha^\mu \bar \alpha^\nu - \alpha^\nu \bar \alpha^\mu \big)\,.
\end{align}
For self-consistency of this solution, $x$ cannot be too close to any of the constituent-instanton locations $z_i$. As long as
\begin{align}\label{eq:faraway}
|x-z_i| \gg \rho_i\,,
\end{align}
the multi-instanton is guaranteed to be of order $\zeta^2$. Otherwise, higher-order corrections need to be taken into account, since for $|x-z_i| \sim \rho_i$ \Eq{eq:AQsmall} can be of order $\zeta^{-1}$; close to their center, constituent-instantons cannot be assumed to be small. Note that the next correction to $A_\mu^{(Q)}$ is of order $\zeta^4$, so \Eq{eq:AQsmall} is accurate to order $\zeta^3$. This observation is relevant for the comparison to the quark zero modes below.

Remarkably, \Eq{eq:AQsmall} corresponds to 't Hooft's multi-instanton solution, but with arbitrary gauge group orientations of the constituent-instantons.
Furthermore, since the single-instanton in singular gauge is given by \cite{Belavin:1975fg}
\begin{align}
A_\mu^{(1)}(x; z,\rho,U) = U \bar\sigma^{\mu\nu} U^\dagger\, \frac{\rho^2}{(x-z)^2}\, \frac{ (x-z)_\nu}{(x-z)^2+\rho^2}\,,
\end{align}
the multi-instanton to order $\zeta^3$ in the SCI limit can be viewed as a superposition of single-instantons,
\begin{align}
A_\mu^{(Q)}(x) = \frac{1}{\xi_0(x) } \sum_{i=1}^{Q} A_\mu^{(1)}(x; z_i,\rho_i,U_i) + \mathcal{O}(\zeta^4)\,.
\end{align}
From \Eq{eq:faraway} follows $(x-z_i)^2+\rho_i^2 \approx (x-z_i)^2$. Here and in the following, this approximate identity is exploited whenever it seems convenient.

For $N_c > 2$, one can simply embed $SU(2)$ into $SU(N_c)$ and use the gauge group orientations $U_i$ from this embedding \cite{Bernard:1977nr}. The corresponding collective coordinates are generated by all global $SU(N_c)$ transformations that yield inequivalent instanton solutions. Since the embedding of $SU(2)$ into $SU(N_c)$ has a stability group $\mathcal{I}_{N_c}$, i.e.\ a subgroup of $SU(N_c)$ that leaves the embedding unchanged, the collective coordinates of the $SU(N_c)$ instanton correspond to the quotient group $SU(N_c)/\mathcal{I}_{N_c}$. Hence, the group integration in \Eq{eq:ZQnQ} is over $SU(N_c)/\mathcal{I}_{N_c}$, with $dU_i$ the corresponding Haar measure.

Since, as proven above, the multi-instanton is a simple superposition of single-instantons in the SCI limit, it has been shown in  
\cite{Brown:1978yj, Bernard:1978ea} that the resulting gauge field contribution to the generating functional factorizes into single-instanton contributions. \Eq{eq:ZQnQ} thus becomes
\begin{align}\label{eq:ZQsmall1}
Z_Q^{\rm SCI}[J] &= \frac{1}{Q!} \int\! \Bigg[\prod_{i=1}^Q d^4z_i\, d\rho_i\, dU_i\, \bar n_1(\rho_i) \Bigg]\, \det{}_0^{(Q)}(J)\,.
\end{align}
The superscript ${}^{\rm SCI}$ indicates that the SCI limit has been applied up to order $\zeta^3$.

Next, the contribution of the quark zero mode determinant needs to be computed. This requires knowledge of the quark zero modes in the SCI limit. They are given by the solutions of the Dirac equation, \Eq{eq:AQDE}, in the background of the multi-instanton in \Eq{eq:AQsmall}. This has first been done in \cite{Pisarski:2019upw} for $Q = 2$. Here, this is generalized to arbitrary $Q$. Using the results of \cite{Corrigan:1978ce, Osborn:1978rn}, the quark zero modes can be obtained directly from the ADHM solution constructed above,
\begin{align}\label{eq:psiQgen}
\psi_{fi}^{(Q)} =  \nu\, \big(N^\dagger C R^{-1} \big)_i\cdot \varphi\,.
\end{align}
This is a left-handed Weyl spinor. $\varphi$ is a constant spinor and $\nu$ a normalization constant. The flavor index, $f = 1,\dots, N_f$, and the `topological charge' index, $i = 1,\dots,Q$, denote the $N_f Q$ quark zero modes. Since the Dirac equation is diagonal in flavor, one simply gets $N_f$ copies of $Q$ manifestly different zero modes. Given the general form of the multi-instanton in \Eq{eq:ADHMA}, it is a straightforward exercise to show that $\psi_{fi}^{(Q)}$ as defined above indeed solves \Eq{eq:AQDE}.

With \Eq{eq:Msmall}, $R^{-1}$ can be computed from \Eq{eq:ADHM1}. $C$ and $N$ are given in Eqs.\ \eq{eq:BC} and \eq{eq:Nsmall}.
The resulting quark zero modes for any $Q$ are
\begin{align}\label{eq:psiQsmall}
\begin{split}
\psi_{fi}^{(Q)}(x) &= \psi_{fi}^{(1)}(x;z_i,\rho_i,U_i)\\
&\quad - \sum_{j \neq i} \mathbb{X}_{ij}(x,z_i)\, \psi_{fj}^{\;(1)}(x;z_j,\rho_j,U_i)\,,
\end{split}
\end{align}
where terms $\sim \rho^n$ with $n \geq 4$ have been dropped and \Eq{eq:faraway} has been used to simplify the expression.
This result is a sum of the single-instanton quark zero modes,
\begin{align}\label{eq:psi1}
\begin{split}
\psi_{fi}^{(1)}(x;z_i,\rho_i,U_i) &= \sqrt{\frac{2}{\pi^2 N_c}} \frac{U_i \rho_i}{\big[(x-z_i)^2+\rho_i^2\big]^{3/2}}\\
&\quad \times \frac{\gamma_\mu (x-z_i)_\mu}{|x-z_i|}\, \varphi_R
\end{split}
\end{align}
and a term which characterizes the overlap between the $Q=1$ zero modes,
\begin{align}\label{eq:ovrlp}
\mathbb{X}_{ij}(x,z_i) = \frac{\rho_i \rho_j |x-z_i|}{[(x-z_i)^2+\rho_i^2]^{3/2}}\,.
\end{align}
The zero mode in \Eq{eq:psiQsmall} is a left-handed Dirac spinor with the right-handed constant spinor
$\varphi_R^{\alpha c}$, where $\alpha$ is a spinor index and $c$ is a color index. For $Q<0$ this is replaced by
a left-handed spinor $\varphi_L^{\alpha c}$. They are normalized to give $\varphi_{R/L}^\dagger \varphi_{R/L} = \mathds{1}_{N_c}$.
The $Q=1$ zero modes are normalized to give
\begin{align}\label{eq:psi1norm}
\int\! d^4 x\, \psi_{fi}^{(1) \dagger}(x;z_i,\rho_i,U_i)\, \psi_{fi}^{(1)}(x;z_i,\rho_i,U_i) = 1\,.
\end{align}
Note that the gauge group orientation of $\psi_{fi}^{(Q)}$ is that of the $i$-th constituent instanton, $U_i$.
\Eq{eq:psiQsmall} is a direct generalization of the result for $Q=2$ in Ref.\ \cite{Pisarski:2019upw}.

The $Q=1$ quark zero mode \eq{eq:psi1} is of order $\zeta$, while the overlap term \eq{eq:ovrlp} is of order $\zeta^2$. Hence, \Eq{eq:psiQsmall} implies that genuine multi-instanton--induced quark zero modes appear at order $\zeta^3$ in the SCI limit. At order $\zeta$, one simply has $\psi_{fi}^{(Q)} = \psi_{fi}^{(1)} + \mathcal{O}(\zeta^3)$. In this case, also the quark zero mode determinant $\det{}_0(J)$ in \Eq{eq:ZQsmall1} factorizes into independent $Q=1$ contributions, which then leads to \Eq{eq:ZQLO}. This gives precise meaning to what has been called the leading order in the SCI limit: it means that the small constituent-instanton expansion is carried out up to order $\zeta^2$. Consequently, no genuine multi-instanton corrections appear at leading order.

The next-to-leading order is $\zeta^3$. As shown here, while the multi-instanton solution, \Eq{eq:AQsmall}, does not change at this order, the quark zero modes do. Hence, the only correction to \Eq{eq:ZQsmall1} at next-to-leading order in the SCI limit stems from the quark zero mode determinant. This correction can be computed from \Eq{eq:psiQsmall}, again, as a direct generalization of the computation in \cite{Pisarski:2019upw}. This is done next.

The quark zero mode determinant in \Eq{eq:ZQsmall1} is
\begin{align}\label{eq:Qdetsmall}
\begin{split}
\det{}_0^{(Q)}(J) = \det \int\! d^4x\, \psi_{fi}^{(Q) \dagger}(x)\, J(x)\, \psi_{gj}^{(Q)}(x)\,,
\end{split}
\end{align}
where the determinant on the right hand side is of the $Q N_f \times Q N_f$ matrix spanned by the zero modes. It is sufficient to look at the contribution of the diagonal elements,
\begin{align}
\begin{split}
&\det{}_0^{(Q)}(J)\Big|_{\rm diag}\\
&\quad = \int\!  \prod_{f=1}^{N_f} \prod_{i=1}^Q\, d^4 x_{fi}\, \psi_{fi}^{(Q) \dagger}(x_{fi})\, J(x_{fi})\, \psi_{fi}^{(Q)}(x_{fi})\,,
\end{split}
\end{align}
the computation of the other contributions is completely equivalent. $x_{fi}$ are the locations of the sources. As discussed above, the integration over all possible instanton locations gives rise to numerous terms which lead, schematically, to \Eq{eq:ZQdec}. For the dilute multi-instanton gas, one only needs to identify the genuine $Q$-instanton contribution $\Delta Z_Q$. With the zero mode solution in \Eq{eq:psiQsmall}, it is straightforward to identify the relevant contributions. The key observation is that every zero mode has a finite overlap with each constituent-instanton. This is reflected in the fact that each $\psi_{fi}^{(Q)}$ depends on all constituent-instanton locations at next-to-leading order in the SCI limit. This is not the case at leading order, where each zero mode only depends on the location of one constituent-instanton. $\Delta Z_Q^{\rm SCI}$ can therefore be extracted from the partition function in \Eq{eq:ZQsmall1} by considering the contributions to the quark zero mode determinant where all zero modes overlap at the same constituent-instanton location. There are $Q$ such contributions. One of them, where all zero modes overlap at $z_1$, reads explicitly
\begin{align}\label{eq:z1ovrlp}
\begin{split}
&\int \Bigg(\prod_{f=1}^{N_f}\prod_{i=1}^Q\, d^4 x_{fi}\Bigg)\\
&\times \prod_{f=1}^{N_f} \Bigg\{ \Big[ \psi_{f1}^{(1) \dagger}(x_{f1};z_1,\rho_1,U_1)\, J(x_{f1})\, \psi_{f1}^{(1)}(x_{f1};z_1,\rho_1,U_1) \Big]\\
&\quad\times \bigg[ \prod_{j=2}^{Q}\psi_{fj}^{(1) \dagger}(x_{fj};z_1,\rho_1,U_j)\, J(x_{fj})\, \psi_{fj}^{(1)}(x_{fj};z_1,\rho_1,U_j) \bigg]\\
&\quad\times \bigg[\prod_{j=2}^{Q} \mathbb{X}_{j1}^2(x_{fj},z_j) \bigg] \Bigg\}\,.
\end{split}
\end{align}
The other $Q-1$ contributions are obtained by replacing $z_1$ with one of the other constituent-instanton locations. Note that the terms in the second and third line of this expression only depend on one location $z_1$. The last term depends on all other instanton locations and therefore quantifies the overlap. 

The overlap term in \eq{eq:z1ovrlp} also depends on the source locations $x_{fj}$. However, if all zero modes $\psi_{fj}^{(Q)}(x_{fj})$ are far away from $z_1$, but closer other $z_j$, then the overlap at these different points would dominate. The resulting contribution is not part of $\Delta Z_Q$, but of $Z_Q^{(q)}$ in \Eq{eq:ZQdec}. Only if all zero modes are closer to one location $z_i$ than to the others (while still being consistent with the SCI limit), i.e.\
\begin{align}
\rho_i \ll |x_{fj}-z_i| \ll |x_{fj}-z_j|\,,
\end{align}  
the quark zero mode determinant generates a genuine $Q$-instanton contribution. In this case, all zero modes for $j \neq i$ are
\begin{align}
\begin{split}
&\psi_{fj}^{(Q)}(x_{f_j})\Big|_{|x_{fj}-z_i| \ll |x_{fj}-z_j|}\\
&\quad \approx - \mathbb{X}_{ji}(z_i,z_j)\, \psi_{fj}^{\;(1)}(x_{fj};z_i,\rho_i,U_j)\,.
\end{split}
\end{align}
Hence, $\mathbb{X}_{j1}^2(x_{fj},z_j)$ in \eq{eq:z1ovrlp} can be replaced by $\mathbb{X}_{j1}^2(z_1,z_j)$, and equivalently for all other contributions

Owing to the different orientations in the gauge group, the zero mode determinant has a nontrivial group structure. To simplify this, one assumes that the source $J$ does not depend on color. In the physical case, the source is identified with the quark mass matrix, $J = \rho M_q$, so this assumption is natural. One can then exploit that the partition function involves the integration over all possible orientations in the gauge group. From the quark determinant, this is of the general form
\begin{align}\label{eq:dUex}
\begin{split}
\int \Bigg( \prod_{i=1}^Q dU_i \Bigg) \prod_{n=1}^{N_f Q}  \varphi_R^{\dagger a_n \alpha_n} U_{i_n}^{\dagger a_n b_n} U_{j_n}^{b_n c_n} \varphi_R^{\alpha_n c_n}\,,
\end{split}
\end{align}
where the indices $i_n$ and $j_n$ are drawn from $2N_f$ copies of the set $\{1,\dots,Q\}$.
The group integrations then result in a combination of products of $\delta^{a_n b_n} \delta^{b_n c_n}$, involving permutations of the indices and $N_c$-dependent coefficients \cite{Creutz:1978ub, Collins_2006}. Due to the normalization of the spinors, $\varphi_R^\dagger \varphi_R = \mathds{1}_{N_c}$, the integration in expression \eq{eq:dUex} leads to fully contracted Kronecker deltas. The gauge group orientations can therefore be rearranged arbitrarily. In particular, \eq{eq:dUex} is identical to
\begin{align}
\begin{split}
&\int \Bigg( \prod_{i=1}^Q dU_i \Bigg) \prod_{n=1}^{N_f Q}  \varphi_R^{\dagger a_n \alpha_n} U_{i_n}^{\dagger a_n b_n} U_{i_n}^{b_n c_n} \varphi_R^{\alpha_n c_n}\\
&\quad= \prod_{n=1}^{N_f Q}  \varphi_R^{\dagger c_n \alpha_n} \varphi_R^{\alpha_n c_n}\,.
\end{split}
\end{align}
Thus, assuming the source $J$ is independent of color, the gauge group integration of the quark zero mode determinant in the SCI limit is trivial.
The second and third line of \eq{eq:z1ovrlp}, as well as the other contributions both from different instanton locations and off-diagonal contributions to the determinant, therefore exactly correspond to the terms that arise from $Q$ copies of the $Q=1$ zero mode determinant.
Defining the $Q=1$ determinant as
\begin{align}
\begin{split}
\det_{0,i}^{\,(1)}(J) = \det\Bigg[ \int\! &d^4 x_{fi}\, \psi_{fi}^{(1) \dagger}(x_{fi};z_i,\rho_i,U_i)\\
&\times J(x_{fi})\, \psi_{gi}^{(1)}(x_{fi};z_i,\rho_i,U_i)\Bigg]\,,
\end{split}
\end{align}
where the determinant is only over flavor in this case, the genuine $Q$-instanton contribution to the partition function in the SCI limit, \Eq{eq:ZQsmall1}, is
\begin{align}\label{eq:DZQ1}
\begin{split}
\Delta Z_Q^{\rm SCI}[J] = \frac{1}{Q!} \int\! &\Bigg[\prod_{i=1}^Q d^4z_i\, d\rho_i\, dU_i\, \bar n_1(\rho_i)\, \det{}_{0,i}^{\,(1)}(J) \Bigg]\\
&\times  \Bigg(\sum_{i=1}^Q \prod_{j\neq i} \mathbb{X}_{ji}^{2 N_f}(z_i,z_j) \Bigg)\,.
\end{split}
\end{align}
To next-to-leading order in the SCI limit, the multi-instanton contribution to the partition function factorizes into single-instanton contributions, multiplied by a nonvanishing overlap term. 

To proceed, it is convenient to specify the source $J$. The canonical choice is the constant diagonal matrix
\begin{align}
J_{ij}^{fg} = \rho_i \delta_{ij} M_q^{fg}\,,
\end{align}  
with the quark mass matrix $M_q$. Due to the normalization of the quark zero modes, \Eq{eq:psi1norm}, the quark zero mode determinant in \Eq{eq:DZQ1} then becomes
\begin{align}\label{eq:det01mass}
\det{}_{0,i}^{\,(1)}(J) = \det{}_{0,i}^{\,(1)}(\rho_i M_q) = \prod_{f=1}^{N_f} \rho_i m_f\,,
\end{align}
where $m_f$ is the mass of the quark flavor $f$. Only the overlap term in \Eq{eq:DZQ1} depends on the instanton locations now. More specifically, it depends on the relative locations $R_{ij}$. Using the explicit form of $\mathbb{X}$ \eq{eq:ovrlp}, the integration over the instanton locations is for $N_f > 1$:
\begin{align}
\begin{split}
&\int\! \Bigg(\prod_{i=1}^Q d^4z_i\Bigg) \Bigg(\sum_{i=1}^Q \prod_{j\neq i} \mathbb{X}_{ji}^{2 N_f}(z_i,z_j) \Bigg)\\
&\quad = \int\! \sum_{i=1}^Q d^4z_i \prod_{j\neq i} d^4R_{ij}\, \Bigg(\frac{\rho_i \rho_j |R_{ij}|}{[R_{ij}^2+\rho_j^2]^{3/2}}\Bigg)^{2 N_f}\\
&\quad = \int\! \sum_{i=1}^Q d^4z_i \prod_{j\neq i} c_{N_f}\, \rho_i^{2 N_f} \rho_j^{4-2 N_f}\,,
\end{split}
\end{align}
with the $N_f$-dependent constant
\begin{align}
c_{N_f} = \frac{\pi^2\, (N_f+1)!\, (2 N_f -1)!}{(3N_f-3)!}\,.
\end{align}
Defining the integrated overlap term as
\begin{align}\label{eq:intovrlp}
I_{Q,N_f}(\{\rho_i\}) = c_{N_f}^{Q-1} \sum_{i=1}^Q \prod_{j\neq i} \rho_i^{2 N_f} \rho_j^{4-2 N_f}\,,
\end{align}
the final result for the genuine $Q$-instanton contribution to the partition function at next-to-leading order in the SCI limit is
\begin{align}\label{eq:DZQfin}
\begin{split}
\Delta Z_Q^{\rm SCI} = \frac{1}{Q!} \int\! d^4z\, \Bigg(\prod_{i=1}^Q d\rho_i\, n_1(\rho_i) \Bigg)\, I_{Q,N_f}(\{\rho_i\})\,,
\end{split}
\end{align}
where the $Q=1$ quark zero mode determinant has been absorbed into the instanton density $n_1(\rho_i) = \bar n_1(\rho_i)\, \det{}_{0,i}^{\,(1)}(\rho_i M_q)$, which is discussed in \App{app:id}.

It is worth emphasizing that \Eq{eq:DZQfin} is the result of a systematic expansion of the exact ADHM solution for the $Q$-instanton. There is only one integration over the instanton location left. It can be interpreted as the integration over the average location of the $Q$-instanton. The relative locations have been integrated out to yield the overlap $I_{Q,N_f}$. This is in accordance with the general discussion that led to \Eq{eq:ZQdec}.

As reviewed in \App{app:id}, the instanton density $n_1(\rho)$ depends non-trivially on $\rho$, so that the integrations over the $\rho_i$ in \Eq{eq:DZQfin} can only be done numerically for each $Q$ and $N_f$. It is therefore not possible to express $\Delta Z_Q$ in closed, analytical form here. For the present purposes, a simple expression of $\Delta Z_Q$ is desirable in order to qualitatively study the $\theta$-dependent free energy $\Delta F(\theta)$ \eq{eq:DF}. A simple estimate is facilitated by the fact that $n_1(\rho)$ has a pronounced peak at an instanton size $\bar \rho$, which can be interpreted as the effective size a constituent-instanton. Using this, the overlap term \eq{eq:intovrlp} can be evaluated directly at $\bar\rho$ for $\Delta Z_Q$, and \Eq{eq:DZQfin} becomes
\begin{align}\label{eq:DZQest}
\begin{split}
\Delta Z_Q^{\rm SCI} \approx \mathcal{V}\ \frac{ \big(\bar c_{N_f} \bar\rho^4 \big)^{(Q-1)}}{(Q-1)!}\, \bar Z_1^Q\,,
\end{split}
\end{align}
where \Eq{eq:Z1} has been used. Since the instanton density has a finite width, $\bar c_{N_f}$ can be quantitatively different from $c_{N_f}$. This result has an intuitive interpretation. The overlap term $I_{Q,N_f}$ accounts for the correlation between $Q$ constituent-instantons. It arises from the short-distance contribution of the integration over their relative distances. The effective volume of a constituent-instanton, in turn, is $\sim \bar \rho^4$. Hence, $\Delta Z_Q^{\rm SCI} \sim \bar \rho^{4(Q-1)}\, \bar Z_1^Q$ reflects the effective geometric overlap of the constituent-instantons that is necessary in order for them to be correlated. This also reflects the short-ranged nature of instanton interactions.

The overlap contribution $\bar c_{N_f} \bar\rho^4$ is in general flavor-, but also temperature-dependent. Here, one final simplification is made by using that, as shown in \App{app:id}, the average instanton size is approximately determined by the renormalization scale parameter $\Lambda$ via $\bar \rho \approx 1/2\Lambda$. This motivates a simple approximation,
\begin{align}
\bar c_{N_f} \bar\rho^4 \approx \Lambda^{-4}\,,
\end{align}
which greatly simplifies the following analysis, as it then only requires knowledge of $\bar Z_1$.
The error of such an estimate can be potentially large, but this is not relevant for the present analysis, where qualitative effects of multi-instantons are explored. Furthermore, while the magnitude of the contribution of the overlap at order $\zeta^3$ in the SCI limit is now known from the analysis above, it is unknown for higher orders. In particular the contributions from instantons themselves, not the quark zero modes, is entirely unknown.
Self-consistency of the SCI expansion at next-to-leading order also restricts the present result to be valid only far away from the constituent-instantons, see \Eq{eq:faraway}. Consistency in the full spacetime region requires higher orders in the SCI limit \cite{Christ:1978jy}, where corrections to \Eq{eq:ZQsmall1} become relevant.
A quantitative analysis therefore requires a more detailed computation of multi-instanton correlations, which is beyond the scope of this work.

\begin{figure}[t]
\centering
\includegraphics[width=.45\textwidth]{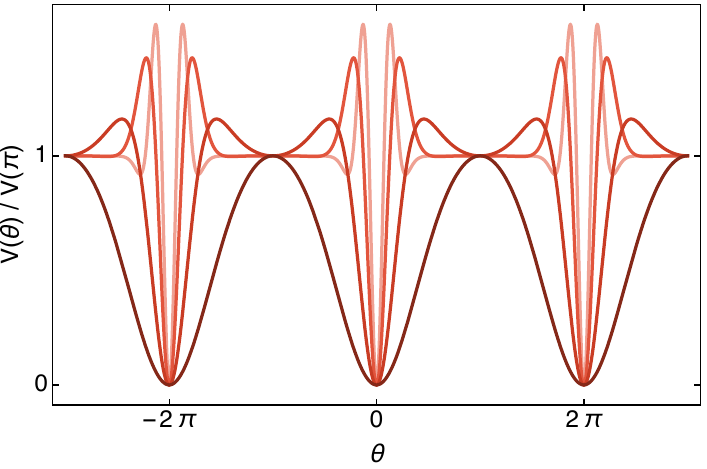}
\caption{The effective potential $V(\theta) = \Delta F^{\rm SCI}(\theta)$, defined in \Eq{eq:FullFree}, for $\widehat Z_1 = 0.1$, 1, 3 and 6 (from dark to light red), normalized by its value at $\theta = \pi$. While the potential retains its $2\pi$-periodicity, strong anharmonicities arise with increasing $\widehat Z_1$.}
\label{fig:pot}
\end{figure}

The free energy is obtained by plugging \Eq{eq:DZQest} into \Eq{eq:DF}. The advantage of using \Eq{eq:DZQest} is that the sum over all topological charges can be carried out analytically. The result is:
\begin{align}\label{eq:FullFree}
\begin{split}
\Delta F^{\rm SCI}(\theta) = 2 \Lambda^4\, \widehat Z_1\Big[e^{\widehat Z_1} - \cos\big(\widehat Z_1 \sin \theta \big)\, e^{\widehat Z_1 \cos \theta} \Big]\,,
\end{split}
\end{align}
with $\widehat Z_1 \equiv \bar Z_1/\Lambda^4$.
For small $\widehat Z_1$, one recovers the well-known result for $Q=1$,
\begin{align}\label{eq:FsmallZ1}
\Delta F^{\rm SCI}(\theta) = 2 \bar Z_1 \big( 1 - \cos\theta \big) + \mathcal{O}(\widehat Z_1^2)\,.
\end{align}
This is expected since the effect of multi-instantons depends on the magnitude of the `tunneling amplitude' $Z_Q$. In the SCI limit, $Q$-instanton corrections are relevant for all $Q$ with $(Q-1) \lesssim \widehat Z_1$. So for $\widehat Z_1 \ll 1$, the free energy is dominated by the conventional $\cos \theta$ behavior, while for $\widehat  Z_1 \gtrsim 1$ sizable corrections to this behavior become relevant. This is illustrated in \Fig{fig:pot}. Such corrections are referred to \emph{anharmonicities}, since for small $\theta$ the free energy is modified with respect to the leading harmonic contribution $\sim \theta^2$. As noted above, this might be misleading if one invokes an acoustics analogy, because multi-instanton contributions are overtones of the single-instanton contribution.

\section{Topological susceptibilities}\label{sec:sus}

The free energy $\Delta F(\theta)$ generates moments of the topological charge distribution, the topological susceptibilities $\chi_n$,
\begin{align}\label{eq:susgen}
\begin{split}
\chi_{2n} &\equiv \frac{\partial^{2n}\Delta F(\theta)}{\partial \theta^{2n}}\Big|_{\theta = 0}\\[1ex]
&= \frac{2 (-1)^{n+1}}{\mathcal{V}} \sum_{Q} Q^{2n}\, \Delta Z_Q\,,
\end{split}
\end{align}
and all odd susceptibilities vanish, $\chi_{2n+1} = 0$.
In the dilute multi-instanton gas, the differences between the susceptibilities $\chi_n$ are due to the different powers of the topological charge, $Q^n$. Hence, if only $Q=1$ is taken into account, all susceptibilities are identical up to the sign,
\begin{align}\label{eq_xhiQ1}
\begin{split}
\chi_{2n}\big|_{Q=1} &= 2 (-1)^{n+1}\, \bar Z_1\,.
\end{split}
\end{align}
In the SCI limit, the topological susceptibilities in \Eq{eq:susgen} resemble moments of the Poisson distribution and can be expressed as
\begin{align}\label{eq:chisci}
\begin{split}
\chi_{2n}^{\rm SCI} = 2 (-1)^{n+1}\Lambda^4\, \widehat Z_1\,e^{\widehat Z_1} \widetilde{T}_{2n}(\widehat Z_1)\,,
\end{split}
\end{align}
where $\widetilde{T}_n(z)$ are polynomials which are related to the Touchard polynomials defined in \App{app:TP}.

It is sometimes convenient to parametrize the $\theta$-dependence of the free energy in terms of deviations from the second susceptibility $\chi_2$,
\begin{align}\label{eq:dfharm}
\Delta F(\theta) = \frac{1}{2} \chi_2 \theta^2 \Bigg[ 1+ \sum_{n=1}^{\infty} b_{2n}(T) \theta^{2n} \Bigg]\,.
\end{align}
Deviations from unity of the expression in the square brackets hence are a measure for the anharmonicity of the $\theta$-dependence of the free energy.
The relations between the anharmonicity coefficients $b_{2n}$ and the susceptibilities are
\begin{align}
b_{2n} = \frac{2}{(2n+2)!} \frac{\chi_{2n+2}}{\chi_{2}}\,.
\end{align}
In case only single-instantons are taken into account, it follows from \Eq{eq_xhiQ1} that the dilute instanton gas predicts constant values for these coefficients,
\begin{align}
b_{2n}\big|_{Q=1} = \frac{2 (-1)^n}{(2n+2)!}\,.
\end{align}
For instantons of any topological charge in the SCI limit, however, one finds
\begin{align}
b_{2n}^{\rm SCI} = \frac{2 (-1)^n}{(2n+2)!} \frac{\widetilde{T}_{2n+2}(\widehat Z_1)}{\widetilde{T}_{2}(\widehat Z_1)}\,.
\end{align}
Thus, an explicit temperature dependence of the coefficients $b_n$ is generated by the effects of higher topological charge.

\begin{figure*}
\centering
\begin{subfigure}[b]{0.32\textwidth}
\includegraphics[width=\textwidth]{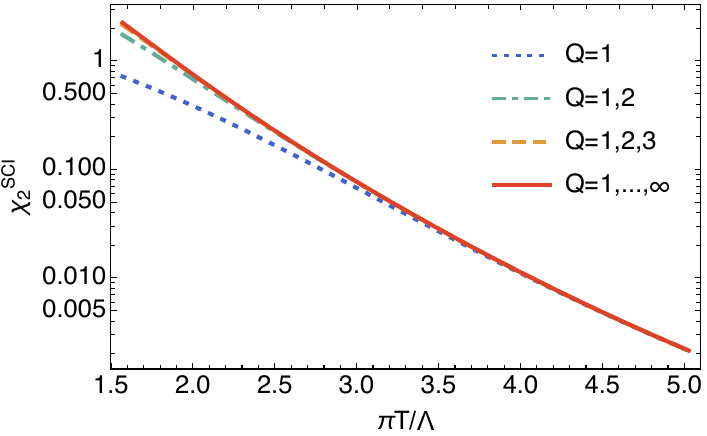}
\caption{}
\end{subfigure}
\hfill
\begin{subfigure}[b]{0.32\textwidth}
\includegraphics[width=\textwidth]{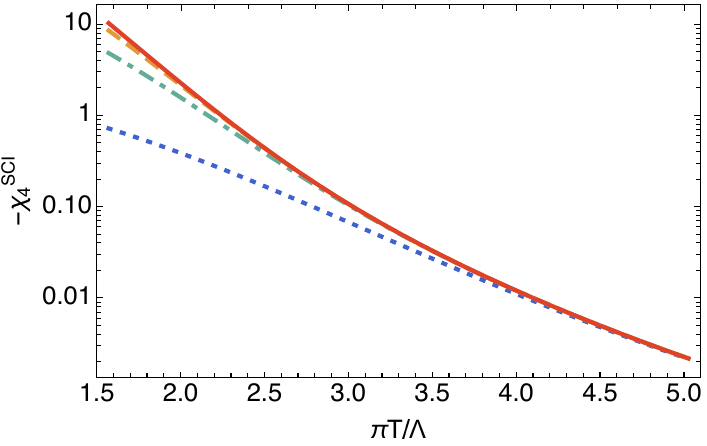}
\caption{}
\end{subfigure}
\hfill
\begin{subfigure}[b]{0.32\textwidth}
\includegraphics[width=\textwidth]{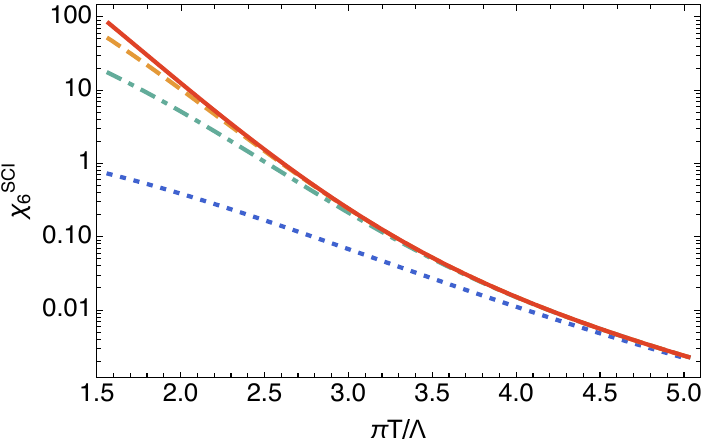}
\caption{}
\end{subfigure}
\caption{Topological susceptibilities $\chi_2^\text{SCI}$ (a), $-\chi_4^\text{SCI}$ (b) and $\chi_6^\text{SCI}$ (c) as functions of temperature in a dilute gas of instantons of various topological charges to next-to-leading order in the limit of small constituent instantons in quenched QCD.}
\label{fig:chis}
\end{figure*}

\subsection{Quenched QCD}\label{sec:YMsus}

Using these results, the topological susceptibilities including the effects of all topological charges can be computed from \Eq{eq:chisci}. 
For the dilute multi-instanton gas in the SCI limit, the size of the multi-instanton contributions to the $\theta$-dependence depends on the size of $\widehat Z_1$. As discussed in \App{app:id}, light quarks lead to a substantial suppression of the instanton density, so in order to study multi-instanton effects within the approximations used here, it is instructive to consider the quenched limit of QCD. In this case, the instanton density of $SU(3)$ Yang-Mills theory is used.

As shown in \App{app:id}, the renormalization group scale is set by the instanton size relative to the renormalization scale parameter $\Lambda$. Thermal corrections to the instanton density only enter through the combination $\pi T \rho$ \cite{Pisarski:1980md, Gross:1980br}. All scales are therefore measured relative to $\Lambda$ here. Since the energy scale of thermal fluctuations is $\sim \pi T$, the combination $\pi T / \Lambda$ sets the relevant thermal scale.

In \Fig{fig:chis} the results for the first nonvanishing susceptibilities $\chi_2^\text{SCI}$, $-\chi_4^\text{SCI}$ and $\chi_6^\text{SCI}$ are shown. The results for the conventional dilute instanton gas, which only accounts for the effect of single-instantons, are compared to the results of dilute gases including also 2- and 3-instantons, as well as all possible $Q$-instantons in the SCI limit. In general, the effect of higher topological charge amplifies the temperature dependence of the susceptibilities. This leads to the general trend that the topological susceptibilities decreases faster with $T$ before they follow the behavior of the dilute single-instanton gas at high temperatures. Such a behavior has been observed on the lattice, see, e.g., \cite{Burger:2018fvb}. Hence, multi-instanton effects provide a microscopic explanation for this. 

Within the range of temperatures considered here, the expansion of the dilute instanton gas in terms of the topological charge $Q$ converges rapidly. This is expected since the instanton density itself is highly suppressed at large temperatures, such that higher powers become less relevant.

\begin{figure*}[t]
\centering
\begin{subfigure}[b]{0.32\textwidth}
\includegraphics[width=\textwidth]{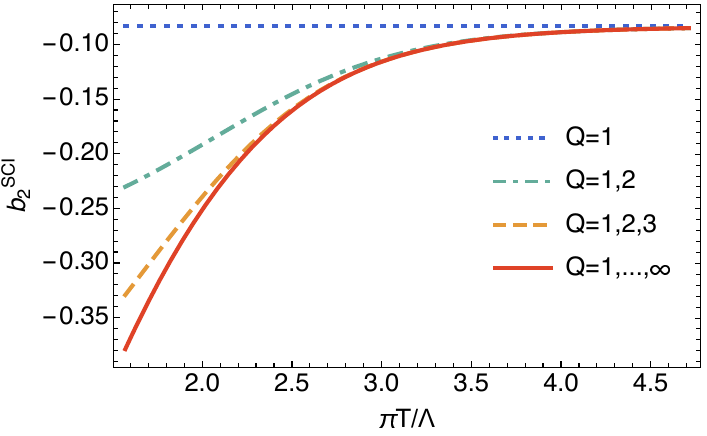}
\caption{}
\end{subfigure}
\hfill
\begin{subfigure}[b]{0.32\textwidth}
\includegraphics[width=\textwidth]{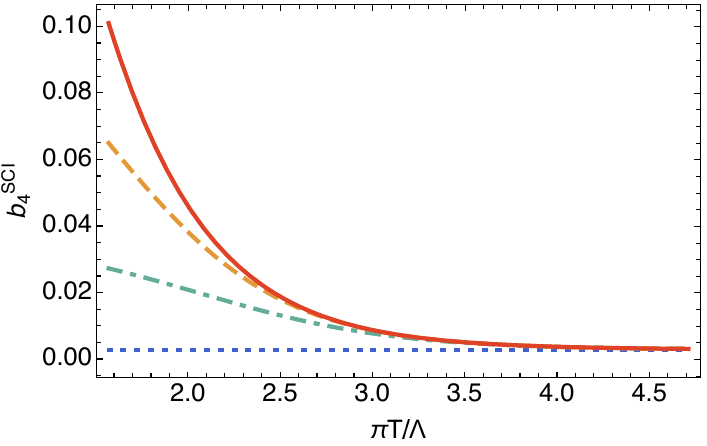}
\caption{}
\end{subfigure}
\hfill
\begin{subfigure}[b]{0.32\textwidth}
\includegraphics[width=\textwidth]{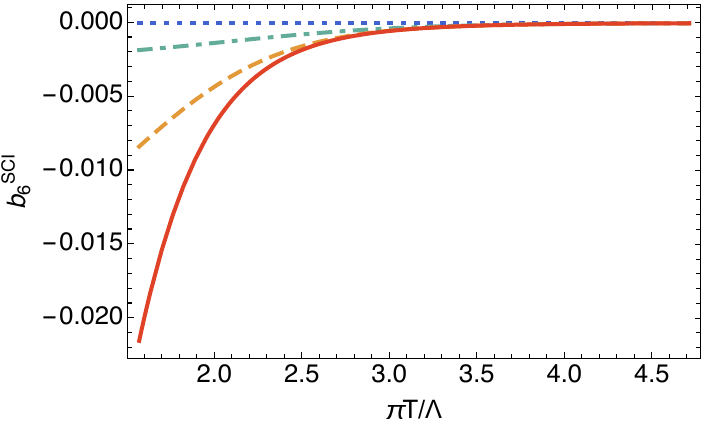}
\caption{}
\end{subfigure}
\caption{Anharmonicity coefficients $b_2^\text{SCI}$ (a), $b_4^\text{SCI}$ (b) and $b_6^\text{SCI}$ (c) as functions of temperature in a dilute gas of instantons of various topological charges to next-to-leading order in the limit of small constituent instantons in quenched QCD.}
\label{fig:bs}
\end{figure*}

To study the effect of anharmonicities induced by multi-instantons, $b_2^\text{SCI}$, $b_4^\text{SCI}$ and $b_6^\text{SCI}$ are shown as functions of temperature in \Fig{fig:bs}. As in \Fig{fig:chis}, the conventional dilute single-instanton gas is compared to the result including any topological charge, as well as with the results of an expansion of the free energy up to $Q = 2$ and $Q=3$. The temperature dependence of the anharmonicity coefficients is solely due to multi-instanton effects. Computations of $b_2$ on the lattice above $T_c$ show indications that, starting from the single-instanton value $-1/12$ at very high temperature, $b_2$ decreases slightly with decreasing temperature for $T \gtrsim 2.5 T_c$, before it starts rising towards small temperatures \cite{Bonati:2013tt, Borsanyi:2015cka, Xiong:2015dya, Lombardo:2020bvn}. As demonstrated here, multi-instanton corrections can explain this behavior qualitatively. However, more precise studies are required to corroborate this on the lattice.

\subsection{QCD}\label{sec:QCDsus}

As discussed after \Eq{eq:FsmallZ1}, the effects of gauge field configurations with higher topological charge in the dilute multi-instanton gas to leading order in the SCI limit depend on the size of the partition function in the presence of a single-instanton, $Z_1$. It is determined by the instanton density $n_1$, which is significantly suppressed in the presence of light quarks, see \App{app:id} and in particular \Fig{fig:z1comp}. Hence, within the approximations used here, the corrections to the leading single-instanton behavior are negligible in QCD for most practical purposes. Explicit numerical calculations show that multi-instanton corrections only become relevant for susceptibilities of very high order. For example, at $\frac{\pi T}{\Lambda} = 1.5$, the effect of 2-instantons on $b_6^\text{SCI}$ is about 0.002\%, while it is about $32\%$ for $b_{20}^\text{SCI}$. In either case, multi-instantons with $Q\geq3$ can be neglected. The present approximations are discussed critically in \Sec{sec:disc}.

\section{Axion cosmology}\label{sec:axi}

As outlined in \Sec{sec:intro}, the physics of axions is sensitive to the topological structure of the QCD vacuum. Higher topological charge effects on the cosmology of axions are explored here.

\subsection{Axion effective potential}\label{sec:axipot}

Having the $\theta$-vacuum as the true vacuum with $\theta$ as a fundamental parameter of the theory begs the question which value is the physical one? Since $\theta \neq 0$ implies $C P$-violation in QCD through $F \tilde F$, one can look corresponding processes in nature. As discussed in \Sec{sec:intro}, measurements of the neutron electric dipole moment put stringent lower bounds on $\theta$, strongly suggesting that its physical value is zero.
The question about the existence and nature of a physical mechanism to enforce this remains to be answered.

Among the possible resolutions of this problem, the PQ mechanism \cite{Peccei:1977hh, Peccei:1977ur} is the one relevant for the present purposes. In this case, the standard model is augmented by an additional global chiral (axial) symmetry, $U(1)_\text{PQ}$, and a complex scalar field (and possibly other fields which are irrelevant here) which is charged under $U(1)_\text{PQ}$ and couples to quarks. This symmetry is spontaneously broken at a scale $f_a$, giving rise to a Goldstone boson, the axion $a(x)$, related to the phase of the complex PQ field. The classical PQ symmetry entails a shift symmetry of $a(x)$. Since $U(1)_\text{PQ}$ is a chiral symmetry, it is anomalous. The only non-derivative interactions of the axion are dictated by the chiral anomaly to be proportional to $F \tilde F$. The axion effective potential is therefore of the same form as the $\theta$-term, and one can define an effective vacuum `angle',
\begin{align}
\bar\theta(x) = f_a \theta + a(x)\,. 
\end{align}
Note that there is also a contribution from the finite quark masses to $\bar\theta(x)$, but this is not relevant here since only $\bar\theta(x)$ itself is of interest for the following discussion. For a more complete discussion, see, e.g., \cite{Peccei:2006as}.

The upshot is that the $\theta$-angle is replaced by a dynamical field $\bar\theta(x)$ (which will also be referred to as the axion for simplicity), so there is a physical value defined by the minimum of the axion effective potential. It follows from the discussion above that the axion effective potential $V$ is identical to the free energy density $\Delta F$. In the dilute multi-instanton gas Eqs.\ \eq{eq:DF} and \eq{eq:FullFree} yield:
\begin{align}\label{eq:axpot}
V(\bar\theta/f_a) &= \frac{2}{\mathcal{V}} \sum_Q \Delta Z_Q \Big[1- \cos(Q \bar\theta/f_a)\Big]\\ \nonumber
&\stackrel{\text{SCI}}{\approx} 2 \Lambda^4\, \widehat Z_1 \Big\{e^{\widehat Z_1} - 2 \cos\big[\widehat Z_1 \sin(\bar\theta/f_a) \big]\, e^{\widehat Z_1 \cos( \bar\theta/f_a)}\Big\}\,.
\end{align}
Hence, the superselection of the $\theta$-parameter is avoided elegantly by effectively promoting it to a dynamical field. The effective potential in the SCI limit is shown for exemplary values for $\widehat Z_1$ in \Fig{fig:pot}. Obviously, the vacuum expectation value is at $\langle\bar\theta\rangle = 0$, which renders the QCD vacuum CP-symmetric. The topological susceptibilities computed in \Sec{sec:sus} can be interpreted directly as the axion mass and its higher order (non-derivative) self-interactions. Thus, the temperature dependence of axion self-interactions is modified by multi-instanton effects.

\subsection{Vacuum realignement}\label{sec:axicos}

The specifics of how the axion couples to the standard model besides the topological sector discussed above are model dependent. There is a class of `invisible' axion models originating from \cite{Kim:1979if, Shifman:1979if, Zhitnitsky:1980tq, Dine:1981rt}, which are consistent with bounds from axion searches \cite{Sikivie:2020zpn}. These models require $U(1)_\text{PQ}$ to be spontaneously broken at a very high energy scale resulting in an axion decay constant of $f_a \gtrsim 10^9\, \text{GeV}$, rendering axions very light and their interactions faint. Furthermore, cold axions can be produced through a field-relaxation mechanism during the evolution of the universe, known as vacuum realignement \cite{Preskill:1982cy, Abbott:1982af, Dine:1982ah, Turner:1985si}. This makes axions viable candidates for dark matter. In the following, the possible implications of higher topological charge effects on the production of cold axions are discussed on a qualitative level.

For illustration, the simplest realization of the vacuum realignement mechanism is used. It is assumed that spontaneous PQ symmetry breaking occurs before inflation and that the reheat temperature is smaller than the temperature for PQ symmetry restoration. Note, however, that the qualitative statements made here are more general. If PQ symmetry is spontaneously broken in the very early universe, the resulting axion is essentially massless since the instanton effects that give rise to the axion mass,
\begin{align}\label{eq:maxion}
m_a^2 = \frac{d^2 V(\bar\theta/f_a)}{d \bar\theta^2} \bigg|_{\bar\theta = 0} = f_a^{-2}\, \chi_2\,,
\end{align}
are negligible at $T \sim f_a$ in this regime. Thus, the axion is strongly fluctuating around its vacuum expectation value within the range 
$\bar\theta/f_a \in [-\pi,\pi]$. In a sufficiently small patch in space right before inflation the axion field can be assumed to have a homogeneous value $\bar\theta_0$. Due to inflation, such a patch is blown up in size and it is possible to have a single homogeneous value $\bar\theta_0$ for the axion within our causal horizon. Furthermore, inflation dilutes all relics from the PQ phase transition, such as topological defects, away. In the simplest realization of the vacuum realignment mechanism, one assumes that we live in one such domain. So it is assumed that the axion is homogeneous throughout the whole universe. Since fluctuations are redshifted away, it can be treated as a classical field. Thus, starting from the random initial value $\bar\theta_0$, called the misalignment angle, the axion evolves in time according to the classical equations of motion. For a more detailed discussion, see, e.g.\ \cite{Sikivie:2006ni, Marsh:2015xka}.

Within the standard model of cosmology, a homogeneous, isotropic, expanding universe with vanishing curvature is assumed. This is described by the Friedmann-Lemaitre-Robertson-Walker metric
\begin{align}
g_{\mu\nu} = \text{diag}\big(1,-a^2(t),-a^2(t),-a^2(t) \big)\,,
\end{align}
with the scale parameter $a(t)$ (not to be confused with the axion). The Hubble parameter is $H(t) =\frac{1}{a(t)} \frac{d a(t)}{dt}$. The axion field described by the classical action
\begin{align}
S_a[\bar\theta] = \int\! d^4x\, \sqrt{- g}\; \bigg[ \frac{1}{2} g^{\mu\nu}\partial_\mu \bar \theta \partial_\nu \bar \theta - V(\bar \theta/f_a) \bigg]\,,
\end{align}
has energy-momentum
\begin{align}
T_{\mu\nu} = \partial_\mu \bar \theta \partial_\nu \bar \theta - g_{\mu\nu}\, \bigg[ \frac{1}{2} g^{\alpha\beta}\partial_\alpha \bar \theta \partial_\beta \bar \theta - V(\bar \theta/f_a) \bigg]\,.
\end{align}
From this, one infers the time evolution of the homogeneous axion field according to its classical equation of motion,
\begin{align}\label{eq:ate}
\frac{d^2\bar\theta}{dt^2} + 3 H \frac{d\bar\theta}{dt} + \frac{d V(\bar\theta/f_a)}{d \bar\theta} = 0\,,
\end{align}
and the energy density of the axion,
\begin{align}\label{eq:ade}
\rho_a \equiv T_{00} = \frac{1}{2} \bigg(\frac{d \bar\theta}{dt}\bigg)^2 + V(\bar\theta/f_a)\,.
\end{align}
Strictly speaking, the time evolution in \Eq{eq:ate} is incomplete. There is an additional Friedmann equation determining the Hubble parameter, which depends on the energy density of the axion. For the timescales relevant here, the universe is to a good approximation radiation-dominated and the axions do not spoil that. Their energy density is negligible compared to the contributions of radiation to the Hubble parameter.

The time evolution of the axion field in \Eq{eq:ate} has a very simple heuristic interpretation if one assumes that anharmonicities in the axion effective potential are small. In this case one has $\frac{d V(\bar\theta/f_a)}{d \bar\theta} \approx m_a^2\, \bar\theta$, cf.\ Eqs. \eq{eq:dfharm} and \eq{eq:maxion}. Naively (ignoring the explicit time dependence of $H$ and $m_a$), \Eq{eq:ate} then has the form of a damped harmonic oscillator. Its qualitative behavior is determined by the damping ratio $\zeta = 3 H / 2 m_a$. If the Hubble parameter dominates over the axion mass such that $\zeta >1$, the axion evolution is overdamped. Since the Hubble expansion is much smaller than the Compton wavelength of the axion at early times, the axion decays very slowly from the initial misalignment angle $\bar\theta_0$ towards its vacuum expectation value. At later times, when a substantial axion mass is generated through instanton effects and the Hubble expansion slows down, the system enters the underdamped regime with $\zeta < 1$. The axion then oscillates with decreasing amplitude around its vacuum expectation value $\bar\theta = 0$.

Due to the strong time-dependence of $H$ and $m_a$ this simple picture is merely suggestive. \Eq{eq:ate} is therefore solved numerically. However, the intuition from the damped oscillator helps to anticipate the possible effect that multi-instanton corrections have on the time evolution of the axion. From the axion potential in \Eq{eq:axpot} shown in \Fig{fig:pot} one sees that the anharmonicity indued by multi-instantons can lead to a flattening of the potential around the maxima, or even turn them into local minima. Thus, the `frequency term' in the axion evolution equation \eq{eq:ate}, $\sim V''(\bar\theta/f_a)$, can become significantly smaller than the corresponding result for the effective potential induced by single-instantons. If the initial misalignment $\bar\theta_0$ happens to be in this flattened region, the axion remains frozen for a longer time before it starts a damped oscillation around its vacuum expectation value for $H^2 \lesssim |V''(\bar\theta/f_a)|$. \Fig{fig:axpot} shown an example for the multi-instanton induced axion effective potential at three different times. A similar scenario, but originating from a noncanonical kinetic term of the axions instead of topological field configurations, has been discussed in \cite{Alvarez:2017kar}. 
Note that if $\widehat Z_1$ becomes large enough, the anharmonic corrections in the SCI limit can induce a sign change in $V'(\bar\theta/f_a)$ for intermediate values of $\bar\theta$. In this case it is possible to \emph{increase} $\bar\theta(t)$, before it starts oscillating. This possibility is commented on below.

\subsection{Quenched-QCD axion}\label{sec:YMaxi}

For the numerical solution of \Eq{eq:ate}, it is assumed that the universe is radiation dominated. The Hubble parameter then is
\begin{align}\label{eq:hubble}
H^2(T) = \frac{8 \pi^3}{90}\, g_\star(T)\, \frac{T^4}{m_\text{Pl}^2}\,,
\end{align}
with the Planck mass $m_\text{Pl} \approx 1.22 \times 10^{19}\, \text{GeV}$. For the effective number of radiative degrees of freedom, $g_\star = 100$ is chosen for simplicity, see, e.g., \cite{Wantz:2009it} for a more refined analysis. Converting temperature to time is done by using that radiation domination implies $t(T) = \frac{1}{ 2 H(T)}$. As for the topological susceptibilities, the axion effective potential is computed in the SCI limit in \Eq{eq:axpot}, with $\widehat Z_1$ discussed in \App{app:id}. As before, all scales are measured relative to $\Lambda$; the relevant time scale is
\begin{align}
 \bar t = \frac{\Lambda}{\pi T}\,.
 \end{align}
To highlight the effect of multi-instantons, an initial misalignment angle close to the maximum of the potential, where the flattening of the potential is most pronounced, and a large axion decay constant, in order to push the onset of the underdamped regime to lower temperatures, are chosen. Specifically, $\bar\theta_0/f_a = 3.14$ and $f_a/\Lambda = 4\times10^{16}$ are used. Given that this is only a toy model, the value for the axion decay constant is not physical.

\begin{figure}[t]
\centering
\includegraphics[width=.45\textwidth]{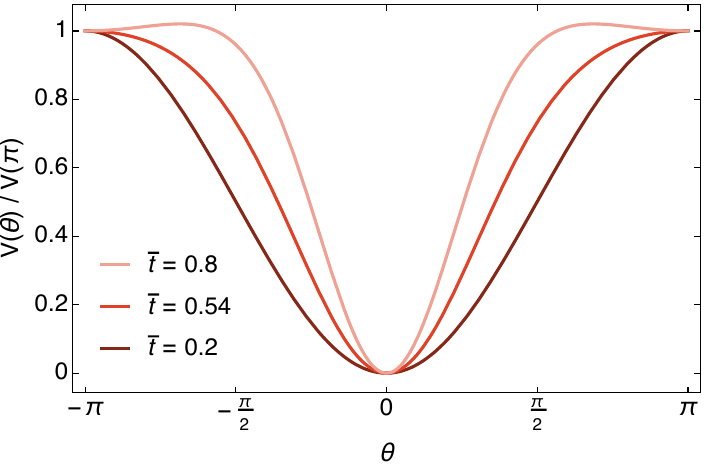}
\caption{Multi-instanton induced axion effective potential for quenched QCD at three different times $\bar t = \Lambda/\pi T$.}
\label{fig:axpot}
\end{figure}

\begin{figure}[t]
\centering
\includegraphics[width=.45\textwidth]{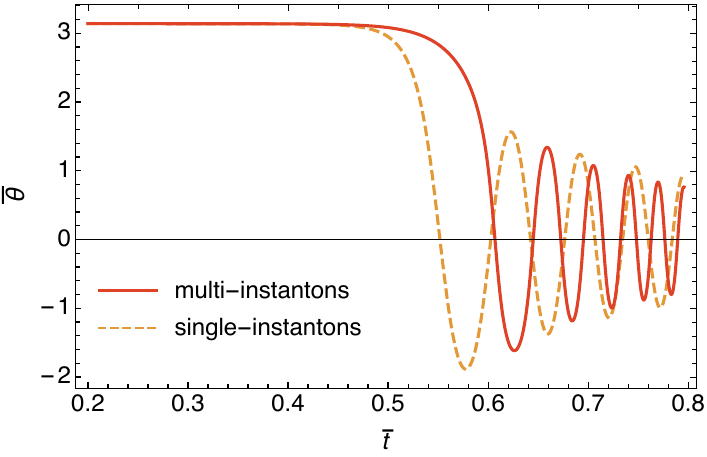}
\caption{Time evolution of the quenched-QCD axion. Multi-instanton effects delay the onset of the oscillating regime and increase the oscillation frequency.}
\label{fig:axevo}
\end{figure}

The time evolution of the axion is shown in \Fig{fig:axevo}. The effect of the multi-instanton induced effective potential of \Eq{eq:axpot} (solid lines) is compared to the single-instanton induced potential $\sim \cos(\bar\theta/f_a)$ (dashed lines). For the present choice of parameters, one sees the qualitative behavior discussed above: the flattening of the potential due to the anharmonicities from multi-instanton corrections results in a longer period of overdamping, where the axion is essentially frozen at the misalignment angle. This can be read-off from \Fig{fig:axpot}, which shows that the multi-instanton induced axion effective potential flattens significantly in the overdamped regime for $\bar t \lesssim 0.54$. In contrast, the single-instanton induced effective potential retains its cosine-shape throughout the whole time evolution. Once the Hubble expansion has slowed down sufficiently, the time evolution is determined by the curvature of the effective potential and the axion oscillates around its vacuum expectation value $\langle \bar \theta \rangle = 0$ with decreasing amplitude.

\begin{figure}[t]
\centering
\includegraphics[width=.45\textwidth]{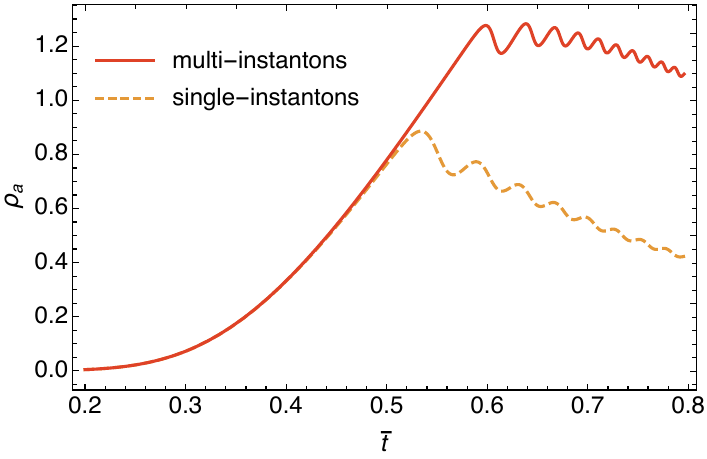}
\caption{Time evolution of the energy density of the quenched-QCD axion. Multi-instanton effects increase the energy density of axions at later times.}
\label{fig:axdens}
\end{figure}

Using this solution, the energy density of the axion can be computed from \Eq{eq:ade}. The result, again for the multi- and single-instanton induced potentials, is shown in \Fig{fig:axdens}. The energy density monotonously rises with time in the overdamped region, where it is almost exclusively due to the potential energy of the axion, and then slowly decreases in the oscillating regime at later times. In addition to the rate of decrease in the oscillating regime, the energy density today crucially depends on how long the axion is frozen in the overdamped regime. Since multi-instanton effects can prolong this phase and delay the start of oscillations, they can increase today's energy density of cold axions. In addition, owing to the steeper potential around the minimum, the axion field oscillates faster in the presence of multi-instanton corrections. Hence, the larger kinetic energy also leads to a larger energy density in the oscillating regime as compared to the single-instanton induced axion potential.
In summary, higher topological charge effects can flatten the axion effective potential and therefore provide a topological mechanism to increase the amount of axion dark matter today.

Following the discussion after \Eq{eq:FsmallZ1}, in \Sec{sec:QCDsus} and \App{app:id}, the substantial suppression of the instanton density in the presence of light quarks leads to also a substantial suppression of multi-instanton effects for the present approximation to the topological structure of the vacuum. Hence, while the mechanism discussed above is still present in QCD, its effect could be negligible. Only axion self-interactions for very high order are modified substantially by multi-instantons in QCD, but their overall magnitude is tiny.

Note that the sign change of $V'$ seen in \Fig{fig:pot} for very large $Z_1$ could lead to an increase of $\bar\theta(t)$ at intermediate times, additionally increasing also the energy density. However, $Z_1$ does never become this large here in the relevant stage of the axion evolution, so this possibility has not been further explored.

\section{Discussion}\label{sec:disc}

Before the results are summarized, a critical discussion of the assumptions and approximations made here is in order. 

First, the actual size of the effects studied here obviously depends on the quantitative impact of field configurations of higher topological charge on the vacuum amplitude. A dilute gas and the SCI limit have been used. A dilute gas can only be valid at weak coupling for very high temperatures. Yet, multi-instanton effects are relevant if the classical suppression $\sim e^{- 8 \pi |Q|/g^2}$ becomes less strong and effective instanton sizes become larger. This is only possible away from the strict weak-coupling limit at lower temperatures. Then, in turn, other effects, such as interactions between (multi-) instantons and (non-perturbative) quantum effects, also become increasingly relevant. Furthermore, it has been argued that the effects in QCD are small because light quarks suppress the instanton density. The dynamical generation of quark mass at lower temperatures might compensate this to some extent. To assess the relative importance of these different effects, a better understanding of the partition function in a $Q$-instanton background, $Z_Q$, beyond the estimates based on the SCI limit used here, is necessary. For this, the overlap between constituent-instantons has to be taken into account more accurately. There is \emph{a priori} no reason to assume that the SCI-limit to a low order is sufficient to accurately describe $Z_Q$, and is therefore the largest source of uncertainty here.

Second, it has been assumed that topological gauge field configurations at large temperatures, i.e.\ well within the deconfined phase, are described by instantons (or rather their finite-temperature cousins, sometimes called calorons \cite{Gross:1980br}).
It has been argued in \cite{Witten:1978bc} that the instanton picture is incompatible at large $N_c$ in the confined phase.
However, neither is $N_c$ large, nor are quarks confined here.
As mentioned already, the assumptions regarding the nature of topological field configurations made here are backed by numerous results for topological susceptibilities by first-principles lattice gauge theory methods. They all show that at large temperatures, $T \gtrsim 2.5 T_c$, the behavior of the susceptibilities with respect to $T$ agrees with the predictions of a dilute instanton gas  \cite{Bonati:2015vqz, Petreczky:2016vrs, Borsanyi:2016ksw, Jahn:2020oqf, Bonati:2013tt, Borsanyi:2015cka, Xiong:2015dya, Berkowitz:2015aua}. Note that this is not contradicting the validity of a dilute multi-instanton gas, since, as shown here, multi-instanton corrections are small at large temperatures and might very well fit within the error bars of state-of-the-art lattice results.

\section{Summary}\label{sec:sum}

It has been shown that gauge field configurations with higher topological charge modify the QCD vacuum. This is reflected in corrections to the conventional dependence on the CP-violating topological $\theta$ parameter.

At large temperatures well within the deconfined phase, the topological structure of QCD can be described by a dilute gas of instantons. Even though multi-instantons with topological charge $Q > 1$ are suppressed in the semi-classical weak-coupling limit, their contributions to the path integral are genuinely different from the single-instanton contribution. Hence, the picture of a dilute instanton gas has been generalized to include instantons of arbitrary topological charge. Note that this is very much in line with the findings in \cite{Pisarski:2019upw}, where it has been shown that there are anomalous quark correlations which are \emph{only} generated by multi-instantons.

In addition to this conceptual result, a key technical result in the present work is the systematic derivation of the multi-instanton contribution to the partition function in the SCI limit for arbitrary topological charges, which has led to \Eq{eq:DZQfin}. This required the explicit expression for the quark zero modes generated in the background of a multi-instanton given in \Eq{eq:psiQsmall}. The computational techniques presented here, which are based on the work in Refs.\ \cite{Pisarski:2019upw, Christ:1978jy, Brown:1978yj, Bernard:1978ea, Corrigan:1978ce, Osborn:1978rn}, pave the way to compute higher-order corrections to the partition function in a multi-instanton background in the SCI limit.

There is a nice acoustics analogy regarding multi-instanton corrections to the $\theta$-dependent free energy of QCD: they give rise to overtones to the fundamental frequency, which is set by single-instantons. 
In the dilute multi-instanton gas in the SCI limit, the $\theta$-dependent free energy can be computed analytically, based on the known results for the single-instanton density. This leads to a modification of the conventional $\cos\theta$-behavior of the free energy.
The resulting $\theta$-dependence of QCD is reflected in the topological susceptibilities $\chi_n$.
If only single-instanton effects are accounted for, all susceptibilities are proportional to the first non-vanishing susceptibility $\chi_2$. Higher topological charge contributions lift this `degeneracy', and amplify the temperature dependence of these susceptibilities towards lower temperatures. This is most clearly seen in the anharmonicity coefficients $b_{2n} \sim \chi_{2n+2}/\chi_2$, which are constant for single-instantons only. Multi-instantons give rise to a characteristic temperature dependence of the anharmonicity coefficients. Strong indications for the behavior of the topological susceptibilities predicted here have been found on the lattice at temperatures above $T_c$. Hence, multi-instanton effects can provide a microscopic explanation for various qualitative features of topological susceptibilities observed on the lattice in the deconfined phase.

An interesting application to showcase higher topological charge effects is axion cosmology. Since the axion effective potential is determined by the $\theta$-dependence of QCD, it is also sensitive to the effects investigated here. Again, to accentuate these effects, a toy universe where axions are coupled to quenched QCD has been considered. The results on the topological susceptibilities directly apply to axion self-interactions. The production of cold axion dark matter via the vacuum realignment mechanism has been studied as an example. Multi-instanton effects can flatten the axion effective potential around its maxima and, as a result, can delay the time where the evolution of the axion switches from the overdamped to the oscillating regime in an expanding, radiation-dominated universe. In addition, the axion oscillates faster. This leads to an overall increase in the energy density of axions at late times, as compared to the case where only single-instantons are taken into account. Hence, higher topological charge effects give rise to a mechanism that increases the amount of axion dark matter.

As discussed in the previous section, on the one hand, the effects studied here become very small if dynamical quarks are taken into account within the present approximations. On the other hand, multi-instanton effects can become relevant only in a regime where semi-classical and dilute approximations begin to break down, at least to leading order.
Thus, a better understanding of the significance of higher topological charge effects requires a more detailed understanding of the impact of topological gauge field configurations on the vacuum amplitude of QCD. This work is a first step in this direction. Two major sources of uncertainty are the unknown higher-order corrections to the vacuum amplitude in a multi-instanton background in the SCI limit, and neglected interactions between (multi-) instantons and anti-instantons.

Higher topological charge effects at lower temperatures, in particular close to and in the confined phase have not been discussed here at all.
As mentioned in the beginning, the reason is that the nature of topological field configurations is unsettled in this case.
A possibility is to study topological effects in `deformed' versions of QCD. One example is Yang-Mills theory on a small circle, where higher topological charge effects can be addressed systematically \cite{Unsal:2012zj}.
At low temperatures, chiral perturbation theory is valid and the $\theta$-dependence can be studied in a controlled manner. This is possible without explicit knowledge of the nature of topological gauge field configurations, since the $\theta$-angle can be rotated to the phase of the quark mass matrix by an axial transformation. The $\theta$-dependence is then encoded in correlation functions of pseudo-Goldstone bosons in the chiral expansion. This way, contributions from all topological charge configurations are taken into account and the free energy also deviates substantially from the simple $\cos \theta$-behavior \cite{Leutwyler:1992yt, Guo:2015oxa, diCortona:2015ldu, Lu:2020rhp}. In this case, it is in principle possible to extract the contributions from different topological charges by means of a Fourier transformation of the free energy with respect to $\theta$ \cite{Leutwyler:1992yt}. However, connecting these results to the present results is impossible since the regions of validity of chiral perturbation theory and dilute instantons do not overlap.

As a final remark, an interesting observation is that multi-instanton effects lead to the possibility of metastable states at $\theta = \pm \pi$, cf.\ Figs.\ \ref{fig:pot} and \ref{fig:axpot}. This could have interesting phenomenological implications due to the possibility of spontaneous $CP$ violation \cite{Dashen:1970et, Witten:1980sp}. 

\vspace{0.2cm} {\bf Acknowledgments --}
The author is grateful to Rob Pisarski
for valuable discussions, comments and collaborations, which inspired this work. It is supported by the U.S. Department of Energy under contract DE-SC0012704.

\begin{appendix}

\section{Instanton density}\label{app:id}

\begin{figure}[b]
\centering
\includegraphics[width=.45\textwidth]{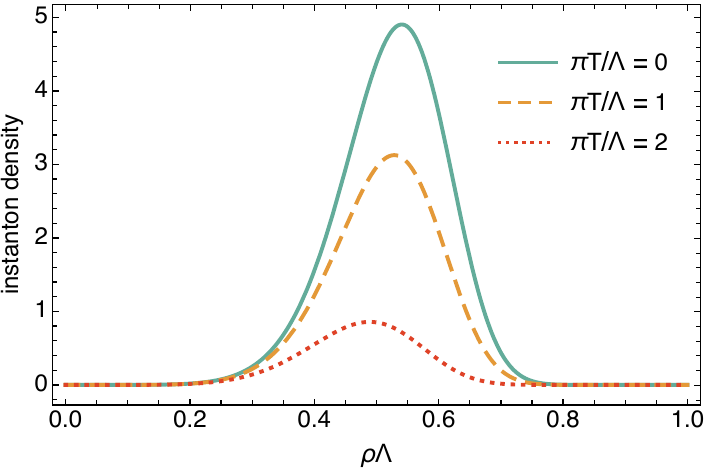}
\caption{Instanton density $n_1(\rho)$ of quenched QCD for various temperatures.}
\label{fig:instden}
\end{figure}

The single-instanton density $n_1$ has been computed in \cite{tHooft:1976rip, tHooft:1976snw, Bernard:1979qt, Morris:1984zi, Pisarski:1980md, Gross:1980br}. In the modified subtraction scheme ($\overline{\rm MS}$), it reads:
\begin{align}\label{eq:instdens}
\begin{split}
n_1(\rho, T) &=  \frac{d_{\overline{\rm MS} } }{\rho^5} \; \left( \frac{8 \pi^2}{g^2}   \right)^6 \; 
\exp\left( - \; \frac{8 \pi^2}{g^2 } \right) \\
&\quad\times \exp\!\bigg[- \frac{2 \pi^2}{g^2}\rho^2 m_D^2(T) - 14 A(\pi T \rho) \bigg]\\
&\quad \times \det{}_0^{(1)}(M_q) \,.
\end{split}
\end{align}
$\det{}_0^{(1)}(M_q)$ is the functional determinant in the space of quark zero-modes and $M_q$ is the quark mass matrix.
The renormalization scheme dependent constant $d_{\overline{\rm MS}}$ is
\begin{align}
d_{\overline{\rm MS}} = \frac{2 e^{5/6}}{\pi^2 (N_c-1)! (N_c-2)!} e^{-1.511374 N_c + 0.2614436 N_f}\,,
\end{align}
$g$ is the running strong coupling. The renormalization group scale is chosen to be determined by the instanton size relative to the scale parameter $\Lambda$, $g(\rho \Lambda)$. The two loop running both in the exponential and the pre-exponential is used \cite{Pisarski:2019upw},
\begin{align}
g^2(x)= \frac{(4 \pi)^2}{\beta_0 \log(x^{-2})} \left(1 - \frac{\beta_1}{\beta_0^2} \frac{\log(\log(x^{-2}))}{\log(x^{-2})} \right) \,,
\end{align}
with $\beta_0 = (11 N_c - 2 N_f)/3$ and $\beta_1 = 34 N_c^2/3 - (13 N_c/3 - 1/N_c ) N_f$. The in-medium corrections depend on the Debye mass at leading order,
\begin{align}
m^2_D(T,\mu) = g^2 \left[\left(\frac{N_c}{3}+\frac{N_f}{6}\right)T^2 + \frac{N_f}{2 \pi^2}\, \mu^2 \right]\,,
\end{align}
and the function
\begin{align}
A(x) = - \frac{1}{12} \log\left(1 + \frac{x^2}{3}\right) + .0129
\left(1 + \frac{0.159}{x^{3/2}}\right)^{-8} \,,
\end{align}
is a numerical parametrization of the temperature-dependent part of the one-loop determinant in the instanton background \cite{Pisarski:1980md, Gross:1980br, Altes:2014bwa}. Note that the instanton density has, at least at the loop order considered here, a large renormalization scheme dependence - the $\overline{\rm MS}$-scheme is just the canonical choice.

The instanton density for the pure $SU(3)$ gauge theory (i.e.\ $N_f = 0$ and $\det{}_0(M_q) = 1$), which is used in the quenched approximation, is shown in \Fig{fig:instden}. First, one clearly sees the suppression of the instanton density due to thermal corrections. Second, the instanton density is peaked about
\begin{align}
\bar\rho \approx \frac{1}{2 \Lambda}\,.
\end{align}
This justifies the use of an effective instanton size in \Sec{sec:ZQSCI}.

\begin{figure}[t]
\centering
\includegraphics[width=.45\textwidth]{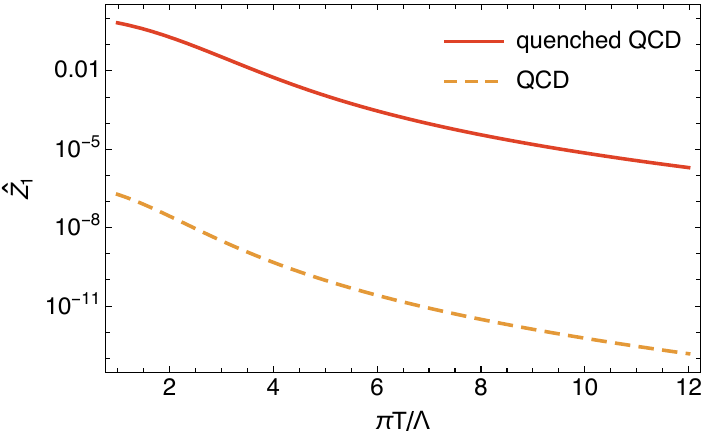}
\caption{Comparison between the single-instanton partition function $\widehat Z_1$ of QCD with and without dynamical quarks.}
\label{fig:z1comp}
\end{figure}

The zero mode determinant of quarks in the background of a single-instanton is given by
\begin{align}
\det{}_0^{(1)}(M_q) = \det \int\! d^4x_{fi}\, \psi^{(1) \dagger}_{fi}(x_{fi}) \, \rho_i \delta_{ij} M_q^{fg}\, \psi^{(1)}_{gj}(x_{fi})\,.
\end{align}
The quark zero modes are defined in \Eq{eq:psi1}. 
Since the quark mass matrix is diagonal in flavor, and one can assume that all instanton sizes are the same in the SCI limit (i.e.\ $\rho_i = \rho$ for all $i$), the determinant simply gives a factor
\begin{align}
\det{}_0(M_q) = \prod_{f=1}^{N_f} \rho m_f\,,
\end{align}
where $m_f$ is the constituent mass of quark flavor $f$. Strictly speaking, this is only valid if $M_q$ can be viewed as a small perturbation of the Dirac operator such that the unperturbed quark-eigenmodes can be used and only the lowest eigenvalue is affected. For a more complete discussion, see \cite{Dunne:2004sx, Dunne:2005te}. For the present purposes, a qualitative discussion is sufficient, though

\Fig{fig:z1comp} shows a comparison between $\widehat Z_1 = \bar Z_1 / \Lambda^4$, defined in \Eq{eq:Z1}, for QCD with and without dynamical quarks. In the former case, four quark flavors with $m_u = m_d = 3\, \text{MeV}$, $m_s = 94\, \text{MeV}$ and $m_c = 1.27\, \text{GeV}$ were chosen. The running of the masses and threshold effects in the running of the strong coupling have been neglected for simplicity. $Z_1$ is suppressed by about seven orders of magnitude if the four lightest quark flavors are taken into account.

\hfill
\section{Touchard polynomials}\label{app:TP}

The evaluation of \Eq{eq:susgen} in the SCI limit involves a summation of the form
\begin{align}
\widetilde{T}_n(z) = z^{-1} e^{-z}  \sum_{Q=1}^\infty Q^n \frac{z^Q}{(Q-1)!}\,.
\end{align}
The polynomials $\widetilde{T}_n$ are directly related to the Touchard polynomials $T_n$ via
\begin{align}
\widetilde{T}_n(z) = \sum_{k=0}^n \begin{pmatrix} n \\ k \end{pmatrix} T_k(z)\,,
\end{align}
and the Touchard polynomials are defined as
\begin{align}
T_k(z) = e^{-z} \sum_{q=0}^\infty q^k\, \frac{z^q}{q!}\,.
\end{align}
They can be written as
\begin{align}
T_n(z) = \sum_{k=0}^n \begin{Bmatrix} n \\ k\end{Bmatrix} z^k\,,
\end{align}
with the Stirling numbers of the second kind
\begin{align}
\begin{Bmatrix} n \\ k\end{Bmatrix} = \frac{1}{k!} \sum_{i=0}^k (-1)^i \begin{pmatrix} k \\ i \end{pmatrix} (k-i)^n\,.
\end{align}
The first few polynomials $\widetilde{T}_n$ are given by
\begin{align}
\nonumber
\widetilde{T}_0(z) &= 1\,,\\ \nonumber
\widetilde{T}_1(z) &= 1+z\,,\\ \nonumber
\widetilde{T}_2(z) &= 1+3 z + z^2\,,\\
\widetilde{T}_3(z) &= 1+7z + 6 z^2 + z^3\,,\\ \nonumber
\widetilde{T}_4(z) &= 1+15z + 25 z^2 + 10 z^3+ z^4\,,\\ \nonumber
\widetilde{T}_5(z) &= 1+31z + 90 z^2 + 65 z^3 + 15 z^4 + z^5\,,\\ \nonumber
\widetilde{T}_6(z) &= 1+63 z + 301 z^2 + 350 z^3 + 140 z^4 + 21 z^5 + z^6\,.
\end{align}


\end{appendix}

\bibliography{instanton}

\end{document}